\documentclass[aps,pre,twocolumn,showpacs,superscriptaddress]{revtex4-1}

\pdfoutput=1
\usepackage[utf8]{inputenc}
\usepackage[T1]{fontenc}
\usepackage{lmodern}
\usepackage{amsmath,amssymb}
\usepackage{graphicx}
\usepackage{float}
\usepackage[usenames,dvipsnames]{xcolor}
\usepackage{tikz}
\usepackage{hyperref}

\newcommand{\eps}{\varepsilon}

\newcommand\etal{{\it et al. }}

\newcommand*\colvec[3][]{
    \begin{pmatrix}\ifx\relax#1\relax\else#1\\\fi#2\\#3\end{pmatrix}
}

\begin{document}

\title{Pattern formation in flocking models: A hydrodynamic description}

\author{Alexandre P. Solon}
\affiliation{Universit\'e Paris Diderot, Sorbonne Paris Cit\'e, MSC, UMR 7057 CNRS, 75205 Paris, France}
\author{Jean-Baptiste Caussin}
\affiliation{Laboratoire de Physique de l'Ecole Normale Sup\'erieure de Lyon, Universit\'e de Lyon, CNRS, 46, all\'ee d'Italie, 69007 Lyon, France}
\author{Denis Bartolo}
\affiliation{Laboratoire de Physique de l'Ecole Normale Sup\'erieure de Lyon, Universit\'e de Lyon, CNRS, 46, all\'ee d'Italie, 69007 Lyon, France}
\author{Hugues Chat\' e}
\affiliation{Service de Physique de l'\'Etat Condens\'e, CNRS UMR 3680, CEA-Saclay, 91191 Gif-sur-Yvette, France}
\affiliation{LPTMC, CNRS UMR 7600, Universit\'e Pierre \& Marie Curie, 75252 Paris, France}
\affiliation{Beijing Computational Science Research Center, Beijing 100094, China}
\author{Julien Tailleur}
\affiliation{Universit\'e Paris Diderot, Sorbonne Paris Cit\'e, MSC, UMR 7057 CNRS, 75205 Paris, France}

\date{\today}

\begin{abstract}
  We study in detail the hydrodynamic theories describing the
  transition to collective motion in polar active matter, exemplified
  by the Vicsek and active Ising models. Using a simple
  phenomenological theory, we show the existence of an infinity of
  propagative solutions, describing both phase and microphase
  separation, that we fully characterize. We also show that the same
  results hold specifically in the hydrodynamic equations derived in
  the literature for the active Ising model and for a simplified
  version of the Vicsek model. We then study numerically the
    linear stability of these solutions. We show that stable ones
    constitute only a small fraction of them, which however includes
    all existing types. We further argue that in practice, a
  coarsening mechanism leads towards phase-separated solutions.
  Finally, we construct the phase diagrams of the hydrodynamic
  equations proposed to qualitatively describe the Vicsek and active
  Ising models and connect our results to the phenomenology of the
  corresponding microscopic models.
\end{abstract}

\maketitle \tableofcontents \makeatletter \let\toc@pre\relax
\let\toc@post\relax \makeatother

\begin{figure*}
  \includegraphics[width=1\textwidth]{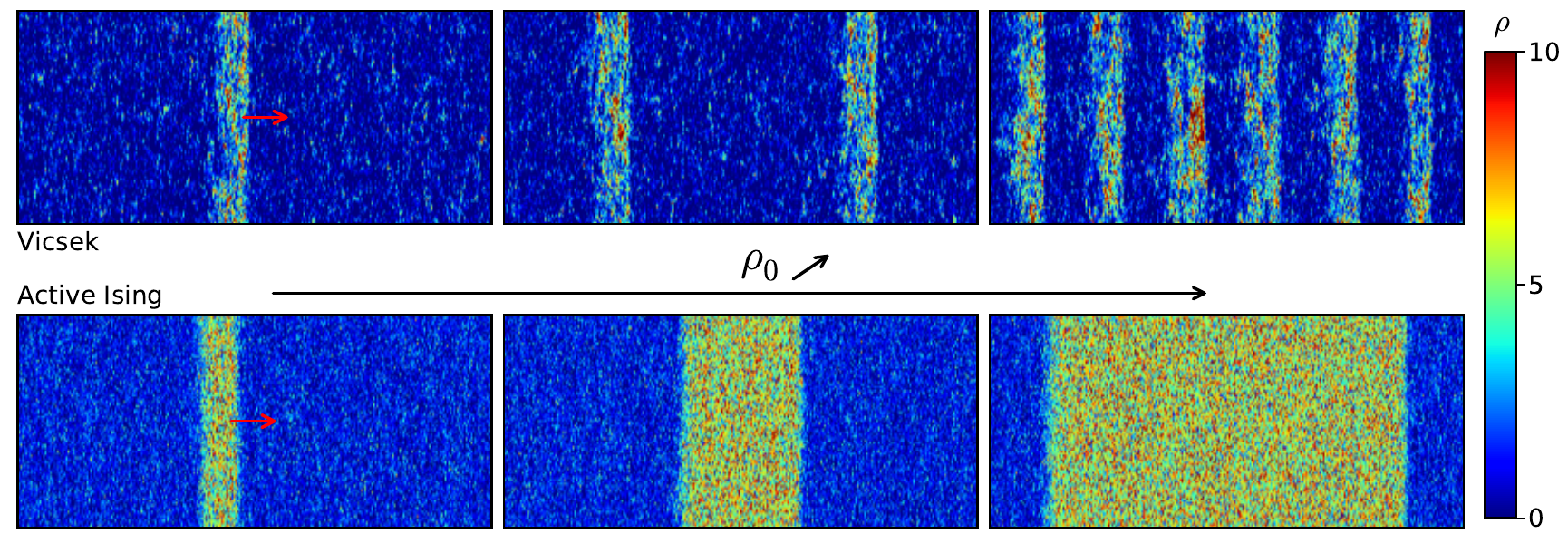}
  \caption{{\bf Top}: Micro-phase separation in the Vicsek
    model. $\eta=0.4$, $v_0=0.5$, $\rho_0=0.83,\,1.05,\,1.93$. {\bf
      Bottom}: Phase separation in the Active Ising model. $D=1$,
    $\eps=0.9$, $\beta{}=1.9$, $\rho_0=1.5,\,2.35,\,4.7$. System sizes
    $800\times 100$. High-density bands propagate as indicated by the
    red arrows on the left snashots.}
  \label{fig:bands_micro}
\end{figure*}

\section{Introduction}

Collective motion is the ability of large groups of motile agents to
move coherently on scales much larger than their individual sizes. It
is encountered at all scales in nature, from macroscopic animal
groups, such as bird flocks, fish schools, or sheep herds, down to the
cellular scale, where the collective migration of
cells~\cite{HakimCells} or bacteria~\cite{PeruaniMB} is commonly
observed. At the subcellular level, in vitro motility assays of actin
filaments~\cite{Schaller2010} or microtubules~\cite{SuminoNature2010}
have shown the spontaneous emergence of large vortices. Collective
motion is also observed in ensemble of man-made motile particles such
as shaken polar grains~\cite{DauchotDisks}, colloidal
rollers~\cite{Rollers}, self-propelled droplets~\cite{Thutupalli} or
assemblies of polymers and molecular
motors~\cite{Schaller2010,SuminoNature2010,Sanchez2012}.  Despite the
differences in their propulsion and interaction mechanisms, these
seemingly very different systems share common macroscopic behaviors
that can be captured by minimal physical models. Of particular
interest is the emergence of {\em directed} collective motion, which
was first addressed in this context in a seminal work by Vicsek and
co-workers~\cite{Vicsek}. The Vicsek model consists in point particles
moving at constant speed and aligning imperfectly with the direction
of motion of their neighbors. When the error on the alignment
interaction is decreased, or the density of particles increased, a
genuine phase transition from a disordered to a symmetry-broken state
is observed. This flocking transition gives rise to an emergent
ordered phase, with true long-range polar order even in 2D, where all
the particles propel on average along the same direction.  Toner and
Tu showed analytically, using a phenomenological fluctuating
  hydrodynamic description, how this ordered state, which would be
forbidden by the Mermin-Wagner theorem at
equilibrium~\cite{MerminWagner}, is stabilized by
self-propulsion~\cite{TT}.  The transition to collective motion in the
Vicsek model has a richer phenomenology than originally thought.  As
first pointed out numerically in~\cite{ChateGregoire}, at the onset of
collective motion, translational symmetry is broken as well. In
periodic simulation boxes, high-density ordered bands of particles
move coherently through a low-density disordered background. The
transition between these bands and the homogeneous disordered profile
is discontinuous, with metastability and hysteresis loops.  These
spatial patterns and the first-order nature of the transition can be
encompassed in a wider framework, which describes the emergence to
collective motion as a liquid-gas phase
separation~\cite{AIM,Microphase}. The travelling bands result from the
phase coexistence between a disordered gas and an ordered polarized
liquid. This framework captures many of the characteristics of the
transition, from the scaling of the order parameter to the shape of
the phase diagram. This phase-separation picture is robust to the very
details of the propulsion and interaction mechanisms. More
specifically, it has also been quantitatively demonstrated in the
active Ising model~\cite{AIM} in which particles can diffuse in a 2d
space but self-propel, and align, only along one axis. However the
specifics of the emergent spatial patterns and the type of phase
separation depend on the symmetry of the orientational degrees of
freedom. While the active Ising model model shows a bulk phase
separation, the Vicsek model is akin to an active XY model and is
associated with a microphase separation where the coherently moving
polar patterns self-organize into smectic structures~\cite{Microphase}
(see Fig.~\ref{fig:bands_micro}).

  In this paper, building on the two prototypical models that are the
  Vicsek model and the active Ising model, we provide a comprehensive
  description of the emergent patterns found at the onset of the
  flocking transition from a hydrodynamic perspective.  We first
  recall the definitions and phenomenologies of these two models in
  Sec.~\ref{sec:micro}. In Sec.~\ref{sec:existence}, we provide a
  simplified hydrodynamic description of the flocking models. In line
  with~\cite{Bertin,Caussin}, we show that these models support
  non-linear propagative solutions whose shape is described using a
  mapping onto the trajectories of point-like particles in
  one-dimensional potentials. Finding such solutions thus reduces to a
  classical mechanics problem with one degree of freedom. For given
  values of all the hydrodynamic coefficients, and hence of all
  underlying microscopic parameters, we find an infinity of solutions,
  describing both phase and microphase separations, that we fully
  characterize. We then show that the same results hold specifically
  for the hydrodynamic equations explicitly derived for the active
  Ising model~\cite{AIM} and for a simplified version of the Vicsek
  model~\cite{Bertin}. Next, we investigate the linear stability
    of these solutions as solutions of the hydrodynamic equations in
  Sec.~\ref{sec:pdestability} and their coarsening dynamics in
  Sec.~\ref{sec:coarsening}. Finally, we provide full phase diagrams
  constructed from the hydrodynamic model in
  Sec.~\ref{sec:phase-diagrams-hydro}. We close by discussing the
  similarities and differences with the phenomenology of the
  agent-based models and conjecture on the role of the hydrodynamic
  noise in the selection of the band patterns.

\section{Phenomenology of microscopic models}
\label{sec:micro}
Let us first briefly recall the phenomenology of the Vicsek and active
Ising models.  They are both based on the same two ingredients:
Self-propulsion and a local alignment rule. The major differences
between the two models are thus the symmetries of the alignment
interaction and of the direction of motion.

\subsection{Vicsek model}
In the Vicsek model~\cite{Vicsek}, $N$ point-like particles, labeled
by an index $i$, move at constant speed $v_0$ on a rectangular plane
with periodic boundary conditions. At each discrete time step $\Delta
t=1$, the headings $\theta_i$ of all particles are updated in parallel
according to
\begin{equation}
  \label{eq:rule_vicsek}
  \theta_i(t+1)=\langle \theta_j(t) \rangle_{j\in \mathcal{N}_i} +\eta\, \xi_i^t
\end{equation}
where $\mathcal{N}_i$ is the disk of unit radius around particle $i$,
$\xi_i^t$ a random angle drawn uniformly in $[-\pi,\pi]$, and $\eta$
sets the level of noise, playing a role akin to that of a temperature
in a ferromagnetic XY model. Then, particles hop along their new
headings: ${\bf r}_i(t+1)= {\bf r}_i(t)+v_0 {\bf e}_i^{t+1}$, where
${\bf e}_i^{t+1}$ is the unit vector pointing in direction given by
$\theta_i({t+1})$.

\subsection{Active Ising model}
In the active Ising model~\cite{AIM}, particles carry a spin
$\pm 1$ and move on a 2D lattice with periodic boundary
conditions. Their dynamics depend on the sign of their spin: A
particle with spin $s$ jumps to the site on its right at rate
$D(1+s\eps)$ and to the site on its left at rate $D(1-s\eps)$, where
$0\le\eps\le 1$ measures the bias on the diffusion. On average, $+1$
particles thus self-propel to the right and $-1$ particles to the left
at a mean velocity $v_0\equiv 2D\eps$. Both types of particles diffuse
symetrically at rate $D$ in the vertical direction.

The alignment interaction is purely local. On a site $i$, a particle
flips its spin $s$ at rate
\begin{equation}
  \label{eq:rule_ising}
  W_i(s\to -s)=\exp\left(-\frac{s}{T}\frac{m_i}{\rho_i}\right)
\end{equation}
where $T$ is a temperature, 
and $m_i$ and $\rho_i$ are the
magnetization and number of particles on site $i$. (An arbitrary number
of particles is allowed on each site since there is no excluded volume
interaction.)
\begin{figure}
  \includegraphics[width=1\columnwidth]{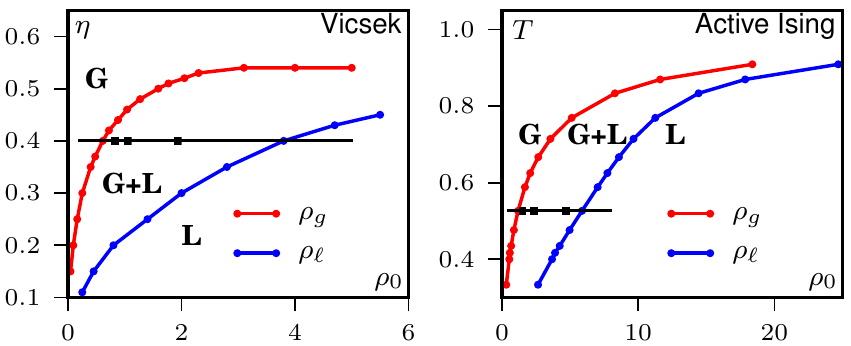}
  \caption{Phase diagrams of the microscopic models. The red and blue
    lines delimit the domain of existence of (micro) phase-separated
    profiles. The black line and squares indicate the position of the
    snapshots shown in Fig.~\ref{fig:bands_micro}. $v_0=0.5$ for the
    Vicsek model, $D=1$ and $\eps=0.9$ for the active Ising model.}
  \label{fig:diagrams_micro}
\end{figure}

\subsection{A liquid-gas phase transition}
\label{sec:LGT}

The phase diagrams in the temperature/noise-density ensemble are shown for
both models in Fig.~\ref{fig:diagrams_micro}, highlighting their
similarity. At high temperature/noise or low density both systems are in a
homogeneous disordered gas state. At low temperature/noise and high density
they are homogeneous and ordered; in these liquid phases, all
particles move in average in the same direction. In the central
region of the phase diagram, inhomogeneous profiles are observed, with
liquid domains moving in a disordered gaseous background.

The phase transitions of both models have all the features of a
liquid-gas transition, exhibiting metastability and hysteresis close
to the transition lines~\cite{ChateGregoire,AIM,Microphase}. The main difference
between the two models lies in the coexistence region: In the active
Ising model, the particles phase separate in a gaseous background and
an ordered liquid band, both of macroscopic sizes~\cite{AIM}. The
coexisting densities depend only on temperature and bias, but not on
the average density; in the coexistence region, increasing the density
at fixed $T,\eps$ thus results in larger and larger liquid domains
whose density remains constant, as shown in
Fig.~\ref{fig:bands_micro}. Conversely, in the Vicsek model, the
  system forms arrays of ordered bands arranged periodically in space
  which have a finite width along their direction of motion: a
  micro-phase separation occurs~\cite{Microphase}. As shown in
Fig.~\ref{fig:bands_micro}, increasing the density at constant noise,
the number of bands increases but their shape does not
change~\cite{Microphase}.

Three types of propagating patterns can thus be observed at phase
coexistence, all shown in Fig.~\ref{fig:bands_micro}: (i) localized
compact excitations, (ii) Smectic microphases, and (iii)
Phase-separated polar liquid domains. In the vicinity of the left
coexistence line, collective motion emerges in the form of localized
compact excitations in both models~\cite{foot1}. At higher density,
phase-separated domains are found in the active Ising model and
periodic ``smectic'' bands in the Vicsek model. Understanding the
emergence of these three types of solutions will be the focus of the
rest of the paper.

\section{Hydrodynamic Equations}
\label{sec:existence}
A lot of attention has been given in the literature to hydrodynamic
equations of flocking models. Two different approaches have been
followed, starting from phenomenological
equations~\cite{TT,Baskaran,Caussin} or deriving explicitly
coarse-grained equations from a microscopic
model~\cite{Bertin,Ihle,Farrell,Pawel,AIM}.  All these equations
describe the dynamics of a conserved density field $\rho(\vec r,t)$
coupled to a non-conserved magnetization field, the latter being a
vector $\vec m(\vec r,t)$ for continuous rotational symmetries, as in
the Vicsek model, or a scalar $m(\vec r,t)$, for discrete symmetries,
as in the active Ising model.

We first introduce in Sec.~\ref{sec:existence_cg} two sets of
hydrodynamic equations derived by coarse-graining microscopic models
which will be discussed in this paper. Then, we turn in
Sec.~\ref{ref:simplerHE} to a simpler set of phenomenological
hydrodynamic equations on which we will establish our general results
in Sec.~\ref{sec:propag}.

\subsection{Coarse-grained hydrodynamic descriptions}
\label{sec:existence_cg}

We first consider the equations proposed by Bertin {\it et al.}
to describe a simplified version of the Vicsek
model~\cite{Bertin}, in which one solely considers binary collisions
between the particles. One can then use, assuming molecular chaos, a
Boltzmann equation formalism to arrive at the following hydrodynamic
equations for the density field and a vectorial magnetization field~\cite{foot3}
\begin{align}
  \label{eq:bertin-hydro-rho}
  \frac{\partial \rho}{\partial t}&=-v_0 \nabla \cdot \vec m\\
  \label{eq:bertin-hydro-m}
  \frac{\partial \vec m}{\partial t}&+\gamma (\vec m\cdot \nabla)\vec m=\nu \nabla^2 \vec m-\frac{v_0}{2}\nabla \rho +\frac{\kappa}{2} \nabla(|\vec m|^2) \nonumber\\
 &\qquad\qquad\qquad -\kappa(\nabla \cdot \vec m)\vec m+(\mu-\zeta |\vec m^2|)\vec m
\end{align}
The mass-conservation equation~\eqref{eq:bertin-hydro-rho} simply
describes the advection of the density by the magnetization
field. Equation~(\ref{eq:bertin-hydro-m}) can be seen as a
Navier-Stokes equation complemented by a Ginzburg-Landau term
$(\mu-\zeta |\vec m|^2)\vec m$, stemming from some underlying
alignment mechanism, and leading to the emergence of a spontaneous
magnetization. Because particles are self-propelled in a given frame
of reference, these equations break Galilean invariance so that one
can have $\gamma\neq 1$ and $\kappa\neq 0$ unlike e.g. in the Navier-Stokes
equation.

In Eqs.~(\ref{eq:bertin-hydro-rho}) and~(\ref{eq:bertin-hydro-m}), to
which we refer to as ``Vicsek hydrodynamic equations'' hereafter, all
the coefficients $\gamma$, $\nu$, $\kappa$, $\mu$ and $\zeta$,
depend on the local density; see~\cite{Bertin} for their exact
  expression.

The second set of equations, which we refer to as ``Ising hydrodynamic
equations'' in the following, has been derived to describe the
large-scale phenomenology of the active Ising model~\cite{AIM}. In
this case, the dynamics of the density field and the scalar
magnetization---corresponding to the Ising symmetry---are given by
\begin{align}
  \label{eq:ising-hydro-rho}
  \frac{\partial \rho}{\partial t}&=D \Delta \rho-v_0\partial_x m\\
  \label{eq:ising-hydro-m}
  \frac{\partial m}{\partial t}&=D \Delta m-v_0\partial_x \rho +2\left(\beta{}-1-\frac{r}{\rho}\right)m -\alpha\frac{m^3}{\rho^2}
\end{align}
where $\beta=1/T$, 
$\alpha$ and $r$ are positive coefficients depending on
$\beta{}$ only and $v_0=2D\eps$. The advection term $v_0 \nabla \vec
m$ of Eq.~(\ref{eq:bertin-hydro-rho}) is here replaced by a partial
derivative $v_0\partial_x m$ because, in the active Ising model, the
density is advected by the magnetization only in the $x$-direction.

\subsection{Phenomenological hydrodynamic equations with constant coefficients}
\label{ref:simplerHE}
Coarse-grained hydrodynamic equations derived from microscopic
  models have the advantage of expressing the macroscopic transport
  coefficients in terms of microscopic quantities (noise,
  self-propulsion speed, etc). However, these possibly complicated
relations may not be relevant to understand the qualitative behavior
of the models. Thus, before discussing the Vicsek and Ising
hydrodynamic equations in Sec.~\ref{sec:CGequations}, we first study
in detail, in Sec.~\ref{sec:propag}, a simpler model
\begin{align}
\label{eq:hydroCC_rho}\partial_t& \rho=-v_0\nabla\cdot \vec m\\
\label{eq:hydroCC_m}\partial_t& \vec m +\xi (\vec m\cdot\nabla) \vec m = D \nabla^2 \vec m -\lambda \nabla\rho +a_2 \vec m -a_4 |\vec m|^2\vec m
\end{align}
where the transport coefficients $v_0$, $\xi$, $D$, $\lambda$ and
$a_4$ are constant. In the following, we refer to these equations as
the ``phenomenological hydrodynamic equations''. This simplified model
is very similar to that first introduced by Toner and Tu from symmetry
considerations~\cite{TT}. However, unlike the original Toner and Tu
model, we keep an explicit density dependence in $a_2$:
$a_2(\rho)=\rho-\varphi_g$, which is essential to account for
inhomogeneous profiles~\cite{Bertin,Baskaran,AIM}. 

The stability criteria of the homogeneous solutions ($\rho(\textbf
r,t)=\rho_0$, $\vec m(\textbf r,t)=\vec m_0$) of
Eqs.~(\ref{eq:hydroCC_rho}-\ref{eq:hydroCC_m}) are readily computed:
\begin{itemize}
\item For $\rho_0<\varphi_g$ ($a_2(\rho_0)<0$) only the disordered solution ($\rho_0$,
  $|\vec m_0|=0$) exists and is stable
\item For $\rho_0>\varphi_g$ ($a_2(\rho_0)>0$) the disordered solution becomes
  unstable and ordered solutions ($\rho_0$,
  $|\vec m_0|=\sqrt{(\rho_0-\varphi_g)/a_4}$) appear.
\item The ordered solutions are linearly stable only when
  $\rho_0>\varphi_\ell=\varphi_g+\frac{1}{4a_4 v_0+2\lambda}$
\end{itemize}
Thus, in the range $\rho_0\in [\varphi_g,\varphi_\ell]$, homogeneous
solutions are linearly unstable. In the language of the liquid-gas
transition, $\varphi_g$ and $\varphi_\ell$ are the gas and liquid
spinodals, between which the homogeneous phases are linearly unstable
and spinodal decomposition takes place. In the next section we address
the existence of heterogenous ordered excitations propagating through
stable disordered (gaseous) backgrounds. This analysis will make it
possible both to identify all the possible flocking patterns and to
further understand the first-order nature of the flocking transition.

\section{Propagative solutions}
\label{sec:propag}
Let us now establish and classify all the inhomogeneous
propagating solutions of
Eqs.~(\ref{eq:hydroCC_rho}-\ref{eq:hydroCC_m}). In order to do
  so, we first recast this problem into a dynamical system
framework in section~\ref{sec:NM}. We then show in
section~\ref{sec:ptot} that three types of propagating solutions exist
with different celerities $c$ and densities of the gaseous background
$\rho_g$. Sections~\ref{sec:stability}, \ref{sec:hopf}
and~\ref{sec:crhoplane} are dedicated to a detailed study of how these
solutions depend on $c$ and $\rho_g$. Section~\ref{sec:fixed-rho0}
shows how, once the average density is fixed, we are left with a
one-parameter family of solutions.  Last, section~\ref{sec:exact} is
devoted to cases where the inhomogeneous profiles can be studied
analytically.

\subsection{Newton Mapping}
\label{sec:NM}
Following~\cite{Bertin}, we look for inhomogeneous polar excitations
invariant along, say the $y$ direction, and which propagate, and/or
relax solely along the $x$ direction. We thus assume, $m_y=0$ and
reduce Eqs.~(\ref{eq:hydroCC_rho}-\ref{eq:hydroCC_m}) to:
\begin{align}
\label{eq:hydroCC_rho1d}\partial_t& \rho=-  v_0\partial_x  m\\
\label{eq:hydroCC_m1d}\partial_t& m +\xi m\partial_x m = D \partial_x^2  m -\lambda \partial_x\rho +(\rho-\varphi_g)  m -a_4  m^3
\end{align}
where we wrote $m=m_x$ to ease the notation. We look for solutions
propagating steadily at a speed $c$. Introducing the position $z=x-ct$
in the frame moving at $c$: $\rho(x,t)=\rho(z)$, $m(x,t)=m(z)$, we
obtain
\begin{align}
\label{eq:hydroCC_rhoz}c&\dot \rho-  v_0 \dot m=0\\
\label{eq:hydroCC_mz}D & \ddot m+(c-\xi m)\dot m-\lambda\dot \rho+(\rho-\varphi_g) m -a_4  m^3=0
\end{align}
where the dots denote derivation with respect to $z$.  Solving
Eq.~(\ref{eq:hydroCC_rhoz}) gives
$\rho(z)=\rho_g+\frac{v_0}{c}m(z)$. If $\rho(z)$ is localized in
space, $\rho_g$ has a simple meaning. Since $\rho(z)=\rho_g$ when
$m(z)=0$, the integration constant $\rho_g$ is the density in the
gaseous phase surrounding the localized polar excitation. We can then
insert the expression of $\rho$ in Eq.~(\ref{eq:hydroCC_mz}) and
obtain the second-order ordinary differential equation
\begin{equation}
  \label{eq:ode}
  D \ddot m+(c-\frac{\lambda v_0}{c}-\xi m)\dot m-(\varphi_g-\rho_g) m+\frac{v_0}{c}m^2 -a_4  m^3=0
\end{equation}
\begin{figure}
  \includegraphics[width=1\columnwidth]{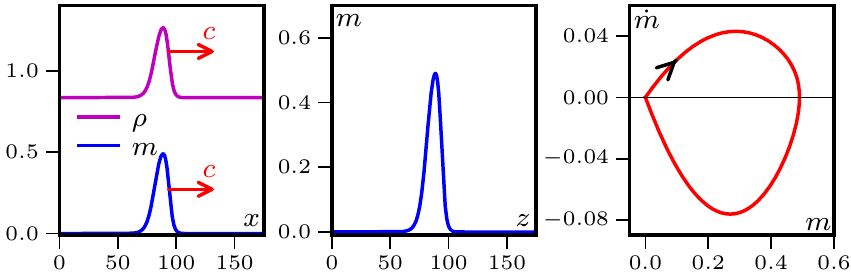}
  \caption{{\bf Left}: Density and magnetization profiles of a
    propagative solution of the hydrodynamic
    Eqs.~(\ref{eq:hydroCC_rho}-\ref{eq:hydroCC_m}). {\bf Center}:
    Magnetization profile in the comoving frame $z=x-ct$ or,
    equivalently, trajectory $m(z)$ of a point particle in the
    spurious time $z$. {\bf Right}: Phase portrait corresponding to
    the trajectory $m(z)$. Parameters:
    $D=v_0=\lambda=\xi=a_4=\varphi_g=1$.}
  \label{fig:first-trajectory}
\end{figure}

Following~\cite{Bertin,Caussin}, we now provide a mechanical
interpretation of Equation~(\ref{eq:ode}) through the well-known
Newton mapping. Rewriting Eq.~(\ref{eq:ode}) as:
\begin{align}
  \label{eq:ode_potential} D \ddot m&=-f(m)\dot m-\frac{dH}{dm} \\
  \label{eq:potential}H(m)&=-(\varphi_g-\rho_g)\frac{m^2}{2}+\frac{v_0}{3c}m^3-\frac{a_4}{4}m^4\\
  \label{eq:friction}f(m)&=c-\frac{\lambda v_0}{c}-\xi m
\end{align}
it is clear that this equation corresponds to the mechanical equation
of motion of a point particle. The position of the particle is $m$,
$z$ is the time variable, $D$ is the mass of the particle, $H(m)$ is
an energy potential and $f(m)$ is a position-dependent friction. The
trajectory $m(z)$ of this fictive particle exactly corresponds to the
shape of the propagative excitations of our hydrodynamic model in the
frame moving at a speed $c$ (see Fig.~\ref{fig:first-trajectory}).

\begin{figure}
  \includegraphics[width=1\columnwidth]{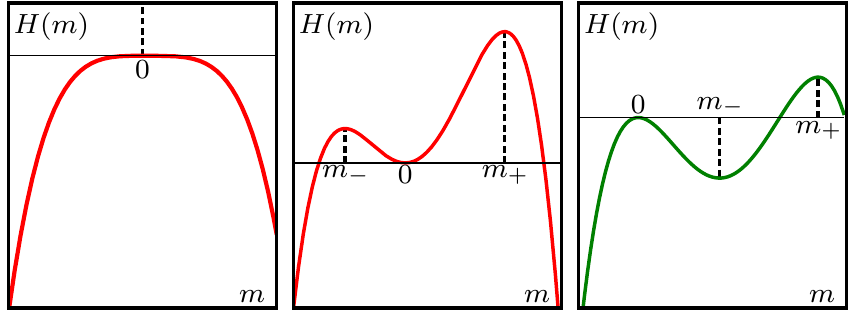}
  \caption{The green potential can give rise to physical (positive,
    non-exploding) solutions while the red ones are ruled out by our
    conditions (C1) (left) and (C2) (center).}
  \label{fig:bad_potential}
\end{figure}

We shall stress that for a given hydrodynamic model,
Eq.~(\ref{eq:ode_potential}) is parametrized by the two unknown
parameters $c$ and $\rho_g$ which {\it a priori} can take  any
value. Each pair ($c$, $\rho_g$) gives different potential $H$ and
friction $f$, and hence different trajectories $m(z)$. We now turn to
the study of these trajectories and of the corresponding admissible
values for the celerity $c$ and the gas density $\rho_g$. 

\begin{figure*}
  \includegraphics[width=1\textwidth]{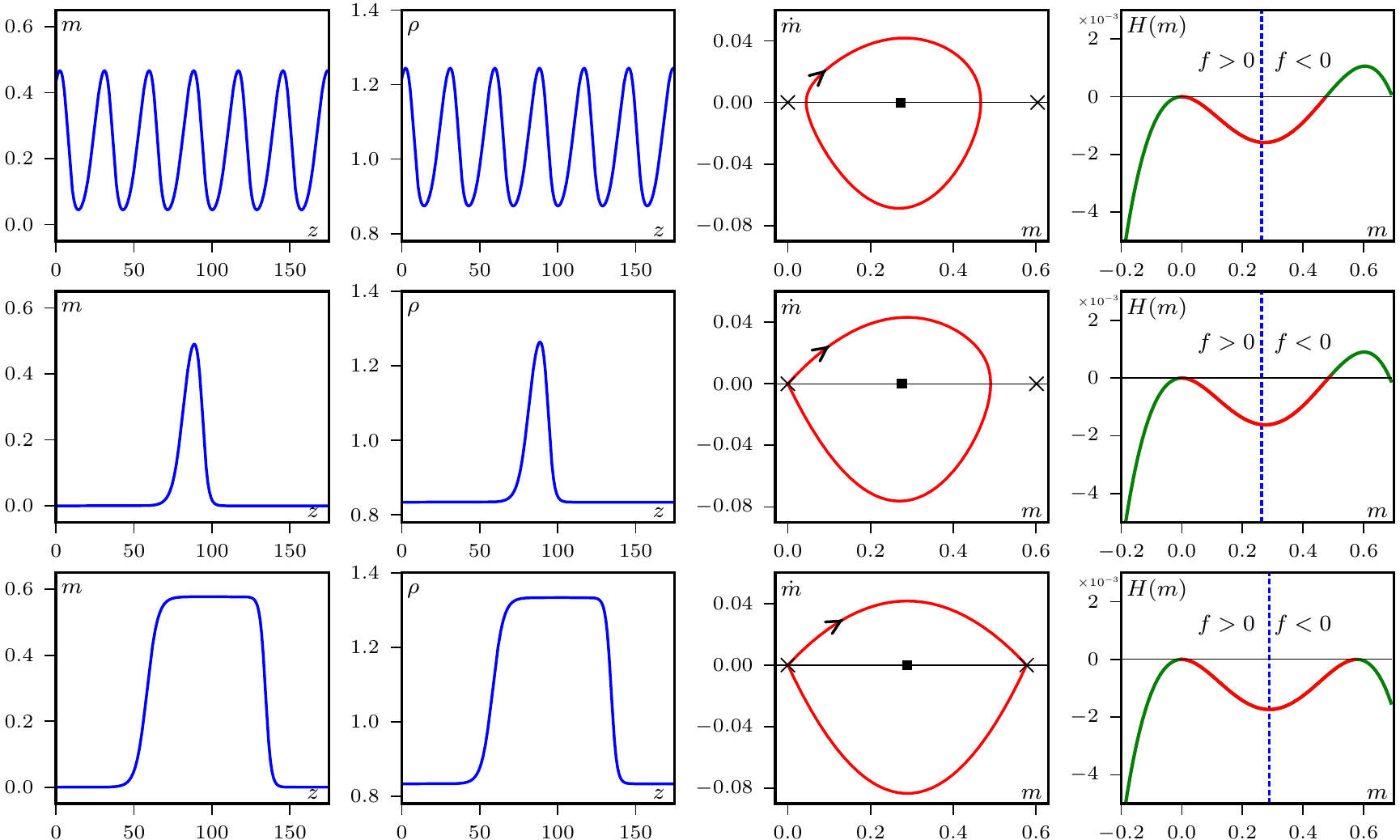}
 \caption{Example of the three types of trajectories. From left to
    right: magnetization and density profiles, phase portrait, and
    potential $H$. {\bf Top row}: Periodic trajectory, $\rho_g=0.835$,
    $c=1.14$. {\bf Center row}: Homoclinic trajectory,
    $\rho_g=0.83412$, $c=1.14$. {\bf Bottom row}: Heteroclinic
    trajectory, $\rho_g=0.83333$, $c=1.1547$. Phase portrait: Crosses
    indicate saddle points at $m=0$ and $m=m_+$. Squares indicate
    stable fixed points at $m=m_-$.  Potentials: The blue dashed line
    indicate where the friction changes sign. The red portion of the
    potential is the one visited by the trajectory. Parameters:
    $D=v_0=\lambda=\xi=a_4=\varphi_g=1$.}
  \label{fig:trajectories}
\end{figure*}

\subsection{Three possible propagating patterns}
\label{sec:ptot}
The original problem of finding all the inhomogeneous propagative
solutions $m(x,t)$, $\rho(x,t)$ of the hydrodynamic equations is now
reduced to finding all the pairs ($c$, $\rho_g$) for which the
corresponding trajectories $m(z)$ exist. Mass conservation,
Eq.~\eqref{eq:hydroCC_rho1d}, imposes the boundary condition
$m(z=-\infty)=m(z=+\infty)$ so that we are looking for solutions
of~\eqref{eq:ode_potential} which are closed in the $(m,\dot m)$
plane. An example of propagative solutions $m(x,t)$, $\rho(x,t)$
together with the corresponding trajectory $m(z)$ and its phase
portraits is shown in Fig.~\ref{fig:first-trajectory}.

To put a first constraint on $\rho_g$ and $c$, let us rule out the
potentials which cannot give such physical solutions. The extrema of
$H$, solutions of $H'(m)=0$, are located at $m=0$ and $m=m_\pm$ with
\begin{equation}
  \label{eq:extrema_H}
  m_\pm=\frac{v_0}{2a_4 c}\left(1\pm \sqrt{1-\frac{4a_4(\varphi_g-\rho_g)c^2}{v_0^2}}\right)
\end{equation}
We can already discard the case where $H'(m)$ has two complex roots
since $H$ then has a single maximum at $m=0$, and all trajectories
wander to $m=\pm \infty$ (see Fig.~\ref{fig:bad_potential},
left). This leads to a first condition on $c$, $\rho_g$:
\begin{equation}\label{eq:C1}\tag{C1}
  (\varphi_g-\rho_g)c^2<a_4 v_0^2
\end{equation}

Without loss of generality we can assume that $c>0$ and look only for
solutions with $m\ge 0$. This rules out the ($c$, $\rho_g$) values for
which $m_-<0$ and $m_+>0$ which give oscillations between negative and
positive values of $m$ (see Fig.~\ref{fig:bad_potential}, central
panel). At the hydrodynamic equation level, such solutions would indeed
correspond to different parts of the profiles moving in opposite
direction. The corresponding condition
\begin{equation}
  \label{eq:C2}\tag{C2}
  \rho_g<\varphi_g
\end{equation}
imposes $0<m_-<m_+$. The potential $H$ then has two maxima, at $m=0$
and $m=m_+$, and one minimum, at $m=m_-$. The typical shape of
potential which gives admissible solutions is shown in
Fig.~\ref{fig:bad_potential} along with examples of potentials ruled
out by conditions~\eqref{eq:C1} and~\eqref{eq:C2}.

>From the admissible shape of the potential $H$, we can now list all possible
trajectories $m(z)$ and the corresponding fields $m(x,t)$,
$\rho(x,t)$:
\begin{itemize}
\item Limit cycles, whose corresponding magnetisation profiles are
  periodic bands, as shown in the first row of
  Fig.~\ref{fig:trajectories}.
\item Homoclinic orbits, that start infinitely close to a maximum of
  $H$, hence spending an arbitrary large time there, before crossing
  twice the potential well in a finite time to finally return to the
  same maximum of $H$ at $z=\infty$. These trajectories correspond
  to isolated solitonic band profiles, as shown in the second row of
  Fig.~\ref{fig:trajectories}.
\item Heteroclinic orbits that spend an arbitrary large time close to
  a first maximum of $H$, cross the potential well in a finite time,
  spend an arbitrary large time close to the second maximum of $H$,
  before returning to the first maximum. These trajectories correspond
  to phase separated profiles. The arbitrary waiting times at the two
  maxima of $H$ then reflect the arbitrary sizes of two
  phase-separated domains (see the third row of
  Fig.~\ref{fig:trajectories}).
\end{itemize}

A third condition on $\rho_g,c$ arises from the non-linear friction
term. Following the classical mechanics analogy, we define an energy
function 
\begin{equation}
  E=\frac1 2 D \dot m^2+H
\end{equation}
Multiplying the equation of motion~(\ref{eq:ode}) by $\dot m$, we get
\begin{equation}
  \label{eq:energy}
  \frac{dE}{dz}=-f(m)\dot m^2
\end{equation}
Energy is injected when $f(m)<0$ and dissipated when $f(m)>0$. On a
closed trajectory, the friction $f$ must thus change sign. Since $f$
is a decreasing function of $m$, this imposes $f(0)>0$ for
trajectories with $m(z)>0$, or equivalently
\begin{equation}\label{eq:C3}\tag{C3}
  c>\sqrt{\lambda v_0}
\end{equation}

The conditions~\eqref{eq:C1}, \eqref{eq:C2} and \eqref{eq:C3} thus
provide loose bounds on the subspace of the ($c, \rho_g$) plane which
contains the three types of trajectories $m(z)$ described above. These
trajectories correspond to the three types of inhomogeneous profiles
seen in the microscopic models \cite{foot2}.  Before studying the
stability and coarsening of these propagative solutions, we first need
to understand precisely how they are organised in the ($c$, $\rho_g$)
plane. In order to do so, we first analyze the phase portrait of the
dynamics~(\ref{eq:ode_potential}). We then study how the phase
portrait evolves when $\rho_g$ and $c$ are varied.

\subsection{Stability of the fixed points}
\label{sec:stability}

The structure of the phase portrait is most easily captured by
locating the fixed points of ~(\ref{eq:ode_potential}) and studying
their stability. We first rewrite (\ref{eq:ode_potential}) as a system
of two first-order differential equations:
\begin{equation}
  \label{eq:ode_system}
  \frac{d}{dz}\colvec{m}{\dot m}=\colvec{\dot m}{-\frac{f(m)}{D}\dot m-\frac{H'(m)}{D}}
\end{equation}
The fixed points are the solutions satisfying $\dot m=0$ and
$H'(m)=0$, {\it i.e.}, the constant solutions extremizing $H$. As
explained before, because of the condition~\eqref{eq:C2}, the extrema of $H$ at
$m=0,m_-,m_+$ are such that $0<m_-<m_+$, so that $0$ and $m_+$ are two
maxima and $m_-$ is a minimum of $H$.

Linearizing the dynamics around one of the fixed points, we define
$m=m_0+\delta m$ with $m_0=0,\, m_-,m_+$, so that $\dot m=\dot{\delta
  m}$ and
\begin{equation}
  \label{eq:ode_system_linear}
  \frac{d}{dz}\colvec{\delta m}{\dot{\delta m}}=
  \begin{pmatrix}
    0 & 1 \\ -H''(m_0)/D & -f(m_0)/D
  \end{pmatrix}
\colvec{\delta m}{\dot{\delta m}}
\end{equation}
The stability of the fixed points is given by the eigenvalues
$\lambda_{1,2}$ of the $2\times 2$ matrix which read
\begin{equation}
  \label{eq:eigenvalues}
  \lambda_{1,2}(m_0)=\frac{-f(m_0)}{2D}\pm \sqrt{\left(\frac{f(m_0)}{2D}\right)^2-\frac{H''(m_0)}{D}}
\end{equation}
%We now discuss the stability of each fixed point.
\begin{itemize}
\item At the maxima $m_0=0$ and $m_0=m_+$ of $H$, $H''(m_0)$ is
  negative and the two eigenvalues are thus real with opposite
  signs. These fixed points are saddle points with one unstable
  direction ($\lambda_1>0$) and one stable direction ($\lambda_2<0$).
\item At the minimum $m_0=m_-$ of $H$, $H''(m_0)$ is positive and the
  real part of the two eigenvalues have the same sign, given by
  $-f(m_-)$. The fixed point is stable when $f(m_-)>0$ and unstable
  when $f(m_-)<0$. Physically, when the friction of the fictive
  particle is negative around $m=m_-$, small perturbations are
  amplified, driving the trajectory away from the fixed
  point. Conversely, a positive friction damps any initial
  perturbation, leading to trajectories converging towards $m_-$.

  When $c$ and $\rho_g$ are such that $f(m_-)=0$, $\lambda_{1,2}$ are
  complex conjugate imaginary numbers. A Hopf bifurcation takes place,
  leading to the apparition of a limit cycle.
\end{itemize}

At the onset of a Hopf bifurcation, a limit cycle appears around the
fixed point whose stability changes. In the following sections we
elucidate how the interplay between the saddle-point and the Hopf
dynamics rules the non-linear dynamics of the fictive particle and
hence the polar-band shape.

\subsection{Hopf bifurcation}
\label{sec:hopf}
Let us first provide a comprehensive characterization of the Hopf
bifurcation. It happens when the real part of $\lambda_{1,2}$
vanishes, {\it i.e.}, when
\begin{equation}
  \label{eq:condition-Hopf}
  f(m_-)=c-\frac{\lambda v_0}{c}-\xi m_-(c,\rho_g)=0
\end{equation}
where $m_-$, which depends on both $c$ and $\rho_g$, is given by
Eq.~(\ref{eq:extrema_H}). Equation~(\ref{eq:condition-Hopf}) is
satisfied on the line
\begin{equation}
  \label{eq:rhogH}
  \rho_g^H(c)=\varphi_g+\frac{(-c^2+v_0\lambda)(-a_4c^2+a_4 v_0\lambda+v_0\xi)}{c^2\xi^2}
\end{equation}
which we call the Hopf transition line.
\begin{figure}
  \begin{tikzpicture}
    \draw (-2.,2) node {\includegraphics{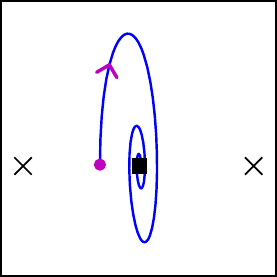}};
    \draw (-2.,-2) node {\includegraphics{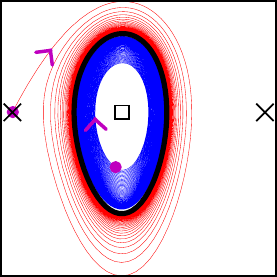}};
    \draw (2.,2) node {\includegraphics{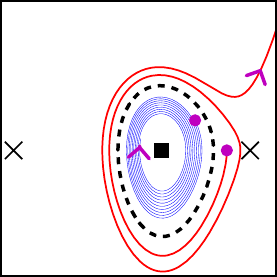}};
    \draw (2.,-2) node {\includegraphics{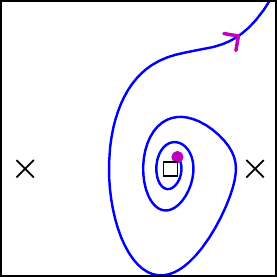}};
    \draw[OliveGreen,dashed,line width=2,->] (-3.5,0)--(3.5,0);
    \draw[OliveGreen,dashed,line width=2,->] (0,-3.5)--(0,3.5);
    \draw (1.8,0.25) node {$\rho_g=\rho_g^H$};
    \draw (-0.2,2) node[rotate=90] {$c=c^*$};
    \draw (3.65,0) node {$c$};
    \draw (0,3.65) node {$\rho_g$};
    \draw (-2,3.65) node {Supercritical};
    \draw (2,3.65) node {Subcritical};
    \draw (-1,0.9) node {\bf I};
    \draw (-1,-3.1) node {\bf II};
    \draw (1.,0.9) node {\bf III};
    \draw (1,-3.1) node {\bf IV};
  \end{tikzpicture}
  \caption{The four types of phase portrait ($m$, $\dot m$) obtained
    in our system when changing $\rho_g$ and $c$. The line
    $\rho_g=\rho_g^H$ is where the Hopf bifurcation takes place. The
    bifurcation is supercritical for $c<c^*$ and subcritical for
    $c>c^*$. The plain (resp. open) black squares denote stable
    (resp. unstable) fixed points at $m=m_-$. The plain (resp. dashed)
    black lines denote stable (resp. unstable) limit cycles. The
    crosses denote the saddle points at $m=0$ and $m=m_+$. The initial
    condition of each trajectory is marked by a magenta disk and the
    direction of ``time'' $z$ indicated by a magenta arrow.}
  \label{fig:schema_hopf}
\end{figure}
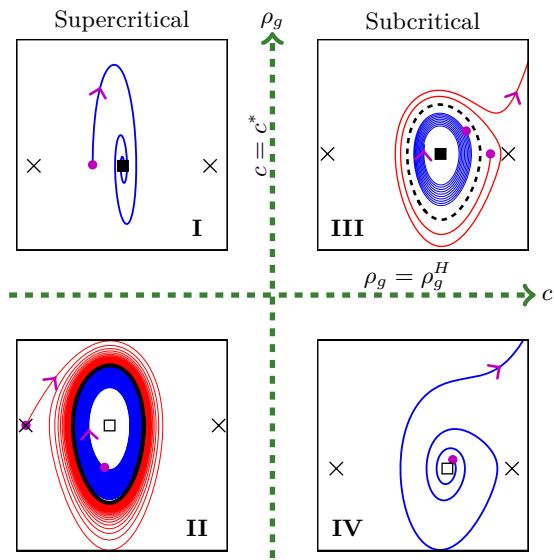

Following standard text books in bifurcation
theory~\cite{Bifurcation}, the type of Hopf bifurcation (super- or
sub-critical) is given by the sign of
\begin{equation}
  \label{eq:delta_Hopf}
  \Delta=\frac{\xi^2}{16\omega D^3}H'''(m_-,\rho_g^H)\frac{\partial m_-}{\partial \rho_g}(\rho_g^H)
\end{equation}
where $\omega=\sqrt{H''(m_-,\rho_g^H)/D}>0$ is the imaginary part of
the eigenvalues at the bifurcation point. Moreover
\begin{equation}
  \label{eq:dmmdrhog}
  \frac{\partial m_-}{\partial \rho_g}=\frac{-1}{\sqrt{\frac{v_0^2}{c^2}-4a_4(\varphi_g-\rho_g)}}
\end{equation}
is always negative because of condition~\eqref{eq:C1}. The sign of
$\Delta$ is thus given by the sign of $H'''(m_-,\rho_g^H)$, which changes
at $c=c^*$ with
\begin{equation}
  \label{eq:cstar}
  c^*=\frac{\sqrt{v_0(3a_4\lambda+\xi)}}{\sqrt{3 a_4}}
\end{equation}
Two different scenarios occur depending on whether $c$ is larger or
smaller than $c^*$.
\begin{itemize}
\item When $c<c^*$, the Hopf bifurcation is supercritical
  ($\Delta>0$). The system branches from a stable fixed point for
  $\rho_g>\rho_g^H$ (case I, Fig.~\ref{fig:schema_hopf}) to a stable
  limit cycle surrounding an unstable fixed point for
  $\rho_g<\rho_g^H$ (case II, Fig.~\ref{fig:schema_hopf}).
\item When $c>c^*$, the Hopf bifurcation is subcritical
  ($\Delta<0$). The system branches from an unstable fixed point when
  $\rho_g<\rho_g^H$ (case IV, Fig.~\ref{fig:schema_hopf}) to an
  unstable limit cycle surrounding a stable fixed point when
  $\rho_g>\rho_g^H$ (case III, Fig.~\ref{fig:schema_hopf}).
\end{itemize}
The organisation of these four typical cases in the $(c,\rho_g)$ plane
is illustrated in Fig.~\ref{fig:schema_hopf}. We thus see
that, when $c<c^*$, limit cycles exist for $\rho_g$ \textit{smaller}
than $\rho_g^H$, whereas when $c>c^*$, they exist for $\rho_g$
\textit{larger} than $\rho_g^H$. The Hopf bifurcation line is thus a
boundary of the domain of existence of periodic propagative solutions
of the hydrodynamic equations. Let us now consider what happens when
we explore the $c,\rho_g$ plane further away from the Hopf bifurcation
line.

\subsection{Structure of the ($c$, $\rho_g$) solution space}
\label{sec:crhoplane}
So far, we have shown that three different types of trajectories
$m(z)$ (periodic, homoclinic and heteroclinic) can be found by varying
the values of $c$, $\rho_g$. The subspace where these physical
solutions can be found was first bounded by the
conditions~\eqref{eq:C1},~\eqref{eq:C2}, and ~\eqref{eq:C3}. In the
previous section, we further found that the Hopf transition line
$\rho_g^H(c)$ given by Eq.~\eqref{eq:rhogH} is the upper boundary for
the admissible values of $\rho_g$ when $c<c^*$ and the lower boundary
when $c>c^*$.

To explore the remaining ($c$, $\rho_g$) space, we numerically
integrated the dynamical system~(\ref{eq:ode_system}) using a
Runge-Kutta scheme of order~4. Starting from different initial
conditions, one easily finds the basins of attraction of the different
solutions. To locate unstable fixed points and limit cycles, we
integrated the dynamics backward in time since they are attractors
when $z\to -\infty$. As $c$ and $\rho_g$ vary, so do the shapes and
sizes of the limit cycles. To quantify these variations, we measured
the ``amplitude'' of a cycle, defined as the difference between the
two extrema of $m(z)$
\begin{equation}\label{eq:deltam}
  \Delta m \equiv \underset{z}{{\rm max}}[m(z)] - \underset{z}{{\rm min}}[m(z)]
\end{equation}

\begin{figure}
  \includegraphics[width=1\columnwidth]{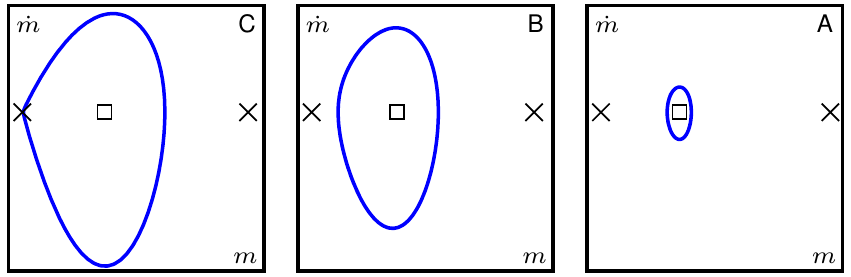}
  \includegraphics[width=1\columnwidth]{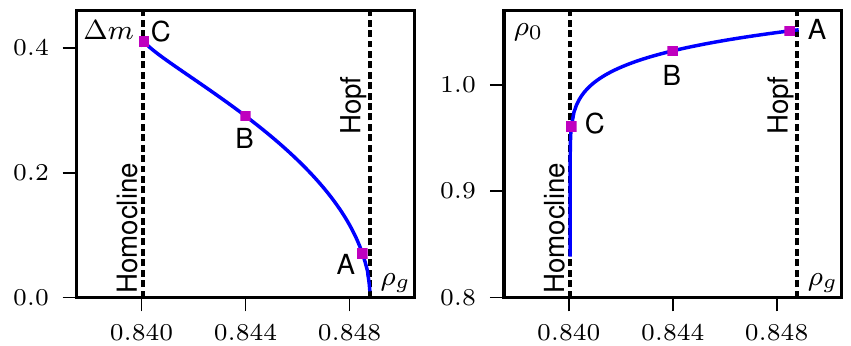}
  \caption{Line at fixed $c=1.12$ in the ($c$, $\rho_g$) space, for
    $c<c^*$.  {\bf Top}: Three phase portraits along the line, from
    the vicinity of the Hopf bifurcation line (A) until the homoclinic
    is found (C) {\bf Bottom-left}: Size of the limit cycle $\Delta m$
    defined in Eq.~\eqref{eq:deltam}. The limit cycle disappears when
    $\Delta m$ is large enough that the orbit reaches $m=0$, where the
    trajectory is homoclinic. {\bf Bottom-right}: Average density
    $\rho_0$ of the solutions.  Parameters:
    $D=v_0=\lambda=\xi=a_4=\varphi_g=1$.}
  \label{fig:line_fixed-c}
\end{figure}

We systematically vary $\rho_g$ at fixed $c$, first focussing on the
case $c<c^*$ where the Hopf bifurcation is supercritical. Decreasing
$\rho_g$, a stable limit cycle of vanishing amplitude appears at
$\rho_g=\rho_g^H$ (Fig.~\ref{fig:line_fixed-c}, panel A). The
amplitude of the cycle then increases as $\rho_g$ decreases
(Fig.~\ref{fig:line_fixed-c}, panel B) until it hits the fixed point
at $m=0$ where the limit cycle becomes an homoclinic trajectory
(Fig.~\ref{fig:line_fixed-c}, panel C). For even lower $\rho_g$ the
particle escapes to $m=-\infty$. The variation of the cycle amplitude
with $\rho_g$ shown in Fig.~\ref{fig:line_fixed-c} can be qualitatively explained. When $\rho_g$ decreases, the distance $m_--m_f$
increases, where $m_f=\frac{1}{\xi}(c-\lambda v_0/c)$ is the value of
$m$ where the friction changes sign, i.e., $f(m_f)=0$. More energy is
thus injected in the system and, to dissipate this energy, the
trajectory need to go closer to $m=0$.

\begin{figure}
  \includegraphics[width=1\columnwidth]{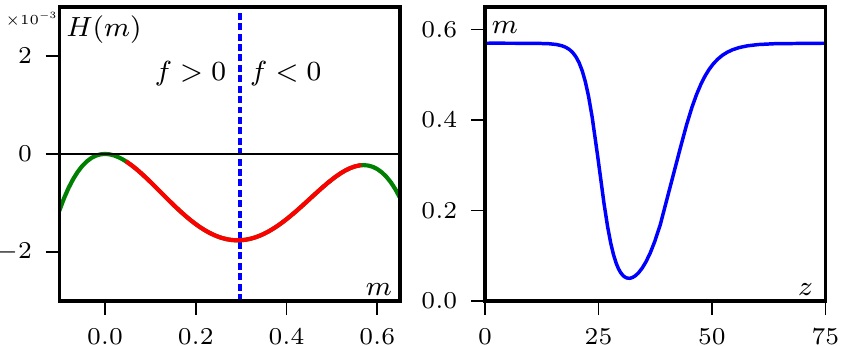}
  \caption{Unstable homoclinic trajectory starting from $m=m_+$ for
    $c=1.16$. Parameters: $D=v_0=\lambda=\xi=a_4=\varphi_g=1$.}
  \label{fig:unstable_homocline}
\end{figure}

A symmetric behavior is observed when $c>c^*$, for the subcritical
Hopf bifurcation. Increasing $\rho_g$, an unstable limit cycle of
vanishing amplitude appears at $\rho_g=\rho_g^H$. The amplitude of the
cycle then increases with $\rho_g$ until the trajectory hits the point
$m=m_+$ where we have an (unstable) homoclinic solution that starts
from $m=m_+$ as shown in
Fig.~\ref{fig:unstable_homocline}. Physically, when increasing
$\rho_g$, $m_f-m_-$ increases so that the friction around the stable
fixed point at $m=m_-$ becomes larger and thus its basin of attraction
(whose boundary is the unstable limit cycle, see
Fig.~\ref{fig:schema_hopf}) becomes larger.

\begin{figure}
  \includegraphics[width=0.8\columnwidth]{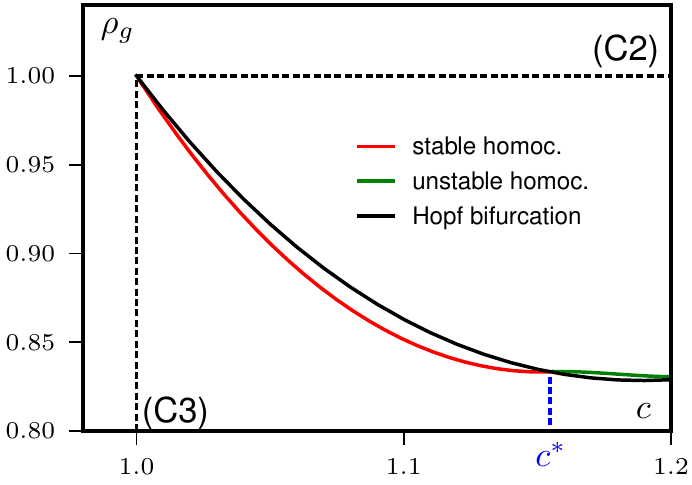}
  \includegraphics[width=0.8\columnwidth]{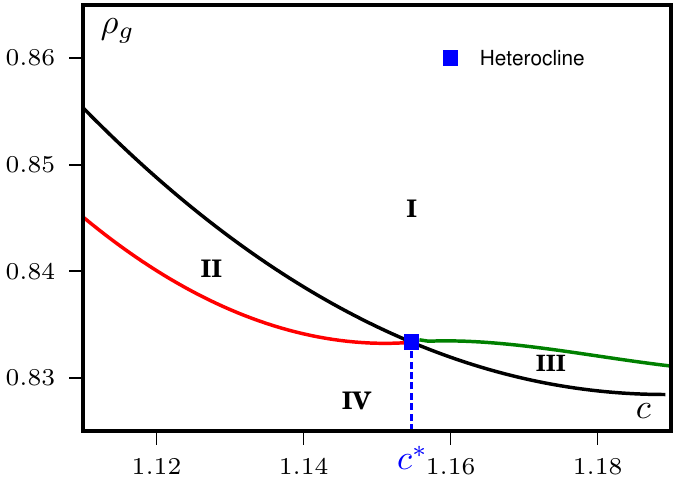}
  \caption{Space of all solutions. A 2-parameter family of periodic
    orbits is found inside the band delimited by the Hopf bifurcation
    and the homoclinic trajectories. The constraints (C2) and (C3) are
    indicated by the black dashed lines. The constraint (C1) lies out
    the range of this plot.  {\bf Bottom}: Zoom of the plot above
    around the point $c=c^*$ where the Hopf bifurcation changes from
    supercritical to subcritical. This is also where the unique
    heteroclinic trajectory is found. The roman numbers refer to
    Fig.~\ref{fig:schema_hopf}, indicating the type of phase portrait
    found in each region. Parameters:
    $D=v_0=\lambda=\xi=a_4=\varphi_g=1$.}
  \label{fig:rhog-c}
\end{figure}

All in all, the central results of this section is that all the
admissible solutions lie in a band delimited by the Hopf bifurcation
line $\rho_g^H(c)$ and a line where the homoclinic trajectories are
found, as shown in Fig.~\ref{fig:rhog-c}. Inside this band there
exists stable non-degenerate limit cycles, corresponding to periodic
propagating profiles. The unique heteroclinic trajectory is located
exactly at $c=c^*$ where the Hopf bifurcation changes from
supercritical to subcritical. We thus observe a 2-parameter family of
periodic solutions, a line of homoclinic trajectories and a unique
heteroclinic trajectory. Going back to the original pattern formation
problem, they correspond to a 2-parameter family of micro-phase
separated profiles, a line of isolated solitonic bands and a unique
phase-separated state where a macroscopic polar liquid domain cruises
through a disordered gas.

\subsection{Working at fixed  average density}
\label{sec:fixed-rho0}
In the microscopic models and the original hydrodynamic equations the
average density $\rho_0$ is a conserved quantity fixed by the initial
condition. On the contrary, when considering the trajectories of the
fictive particle $m(z)$, $\rho_0$ is not \textit{a priori} fixed and
varies between the different solutions. To compute the mean density on
a trajectory $m(z)$ we simply average $\rho(z)=\rho_g+v_0 m(z)/c$ over
time.

As shown in Fig.~\ref{fig:line_fixed-c} (bottom-right), we find that
at fixed $c<c^*$, $\rho_0$ decreases when $\rho_g$ decreases. It
ranges from $\rho_0=\rho_g+\frac{v_0}{c}m_-$ when $\rho_g=\rho_g^H$ to
$\rho_0=\rho_g$ at the homoclinic trajectory where the portion of the
trajectory with $m(z)\neq 0$ becomes negligibly small. Note that, at
the heteroclinic trajectory, $\rho_0$ can take a large range of
values. Since the size of the gas and liquid domains are arbitrary,
the average density can take any value in $[\rho_g^h,\rho_\ell^h]$
where $\rho_g^h$ and $\rho_\ell^h$ are the densities in the gas and
liquid domains respectively.

\begin{figure}
  \includegraphics[width=0.8\columnwidth]{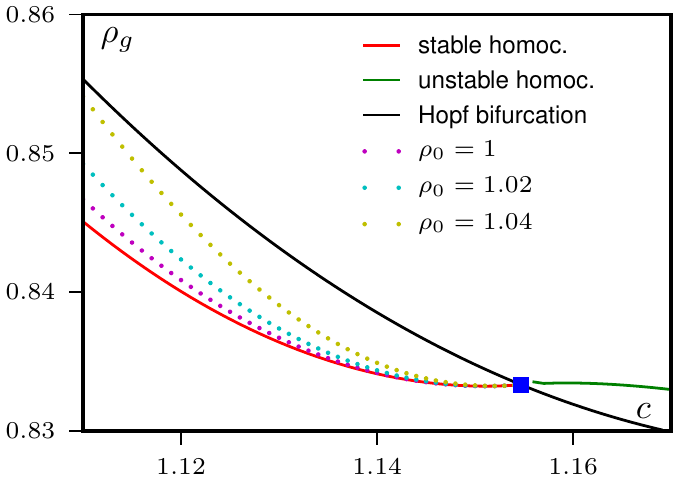}
  \includegraphics[width=0.8\columnwidth]{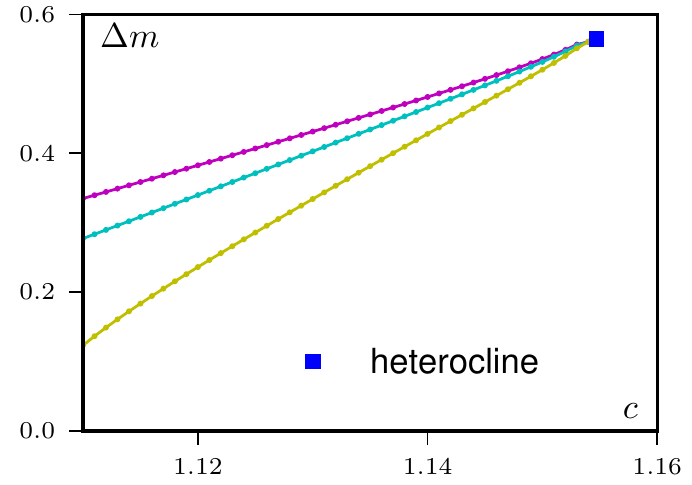}
  \caption{{\bf Top}: Lines of solutions having a fixed average
    density $\rho_0$ in the space of all solutions. {\bf Bottom}: the
    amplitude $\Delta m$ of the cycles, defined in
    Eq.~\eqref{eq:deltam}, along the lines $\rho_0=\text{cst}$
    increases when $c$ increases. Parameters:
    $D=v_0=\lambda=\xi=a_4=\varphi_g=1$.}
  \label{fig:fixed-rho0}
\end{figure}

Fixing $\rho_0$ adds a constraint that selects a line of solutions in
the ($c$, $\rho_g$) space, as shown in Fig.~\ref{fig:fixed-rho0}
(left). For all $\rho_0\in [\rho_g^h,\rho_\ell^h]$ these lines end at
the heteroclinic trajectory. We also observe that, at fixed $\rho_0$,
the closer the trajectories are to the heteroclinic solution, the
larger their amplitude (see Fig.~\ref{fig:fixed-rho0}, right). This
means that along a line $\rho_0=\text{cst}$, the higher the amplitude
the faster the band excitations propagate. This point will turn out to
be crucial when discussing the coarsening dynamics at the hydrodynamic
level in Sec.~\ref{sec:coarsening}.

Until now, we have shown that three different types of possible
trajectories $m(z)$ exist, which correspond to all the propagative
solutions observed in the microscopic models of flying spins. We have
further identified the subset of values of the propagation speed $c$
and the gas density $\rho_g$ for which these solutions exist. We can
now turn to the study of their dynamical stability at the hydrodynamic
equation level. However, we first discuss analytically in the next
section the shape of inhomogeneous solutions.

\subsection{Exact solution for the heterocline}
\label{sec:exact}
There are no general analytic solutions for the propagating
inhomogeneous profiles. However, progress is possible for some
limiting cases. In the following we show that a complete solution for
the heterocline -- its position in the ($c$, $\rho_g$) plane and its
shape -- can be determined exactly~\cite{AIM}. In~\ref{sec:smallD}, we
then show that, although exact solutions are not available, progress
regarding the shape of the homoclinic solutions is achievable in the
small $D$ limit.

\begin{figure}
  \includegraphics[width=0.75\columnwidth]{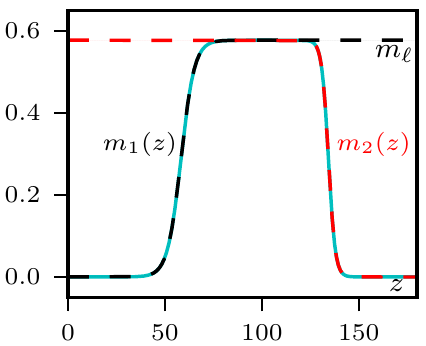}
  \caption{Comparison between the exact solution for the heterocline
    (dashed lines) and the result from numerical integration of
    Eq.~(\ref{eq:ode_system}) (blue line). Parameters:
    $D=v_0=\lambda=\xi=a_4=\varphi_g=1$.}
  \label{fig:heterocline}
\end{figure}

To compute the shape of the heterocline, let us start from the ansatz
\begin{equation}
  \label{eq:ansatz}
  m_{1,2}(z)=\frac{m_\ell}{2}\Big[1+\tanh\big(k_{1,2}(z-z_{1,2})\big)\Big]
\end{equation}
Each of $m_1(z)$ and $m_2(z)$ describe an interface centered around
$z=z_{1,2}$ between a disordered phase with $m=0$ and an ordered phase
with $m=m_\ell$. The complete heteroclinic trajectory then consists of
two fronts glued together: An ascending front $m_1(z)$ with $k_1>0$
and a descending front $m_2(z)$ with $k_2<0$, with $z_2 \gg z_1$ (see
Fig.~\ref{fig:heterocline}); Being part of the same profile, the two
fronts share the same celerity $c$ and density $\rho_g$.

Moreover we know that $m_\ell$ must be located at the second maximum
of $H$ so that
\begin{equation}
  \label{eq:ml}
  m_\ell=m_+=\frac{v_0}{2a_4 c}\left(1+ \sqrt{1-\frac{4a_4(\varphi_g-\rho_g)c^2}{v_0^2}}\right)
\end{equation}
Plugging $m_1$ and $m_2$ in Eq.~(\ref{eq:ode}) and replacing $m_\ell$
by its expression, one obtains for each of the fronts
\begin{equation}
  \label{eq:front-tanh}
  g(c,\rho_g,k_{1,2})+h(c,\rho_g,k_{1,2})\tanh\left(k_{1,2}(z-z_{1,2})\right)=0
\end{equation}
where $g$ and $h$ are complicated functions that we omit for
conciseness. Eq.~(\ref{eq:front-tanh}) can be true only if $g$ and $h$
vanish independently, for both $k_1$ and $k_2$.

We can express $k_1$ and $k_2$ as functions of $c$ and $\rho_g$ by
linearizing the ansatz~(\ref{eq:ansatz}) around $m=0$. When
$k_{1,2}(z-z_{1,2})\to -\infty$, one has $m_{1,2}\sim
\exp(2k_{1,2}(z-z_{1,2}))$ so that we can identify $k_{1,2}$ with the
two eigenvalues of the linear stability analysis
Eq.~(\ref{eq:eigenvalues}). The ascending front is associated with the
unstable direction $k_1=\lambda_1/2$ and the descending front with the
stable direction $k_2=\lambda_2/2$.

Replacing $k_{1,2}$ by their values in Eq.~(\ref{eq:front-tanh}), we
have four equations for the two unknowns $c$ and $\rho_g$. After some
algebra, one obtains a unique solution ($c^h$, $\rho_g^h$) with
\begin{align}
  \label{eq:sol-heterocline}
  c^h&=c^*=\frac{\sqrt{v_0(3a_4\lambda+\xi)}}{\sqrt{3 a_4}} \\
  \rho_g^h&=\varphi_g-\frac{2 v_0}{9 a_4\lambda+3\xi}
\end{align}

This gives us the magnetization $m_\ell$ and the the fronts steepness
$k_{1,2}$ as
\begin{align}
  \label{eq:sol2-heterocline}
  m_\ell&=\frac{2 v_0}{\sqrt{3a_4 v_0(3 a_4\lambda+\xi)}} \\
  k_1&=\frac{\sqrt{v_0(8 a_4 D+\xi^2)}-\sqrt{v_0}\xi}{4 D\sqrt{3 a_4(3 a_4\lambda+\xi)}}\\
  k_2&=\frac{-\sqrt{v_0(8 a_4 D+\xi^2)}-\sqrt{v_0}\xi}{4 D\sqrt{3 a_4(3 a_4\lambda+\xi)}}
\end{align}

In Fig.~\ref{fig:heterocline}, we show that this solution matches
exactly the heteroclinic orbit found by numerical integration of
Eq.~(\ref{eq:ode_system}).

\section{Back to the Vicsek-like and the active Ising models: non-linear solutions of the hydrodynamic equations}
\label{sec:CGequations}

In Sec.~\ref{sec:existence} and~\ref{sec:propag} we consider
phenomenological hydrodynamic equations and assumed the simplest possible
dependences of their coefficients with density. Here we extend our
study to the more realistic hydrodynamic equations presented in
Sec~\ref{sec:existence_cg}. We first consider in
Sec.~\ref{sec:hydrovic} the Vicsek hydrodynamic equations before
turning to the Ising hydrodynamic equations in
Sec.~\ref{sec:hydroising}. For sake of completeness, we also consider
in Appendix~\ref{sec:a4ofrho} a more general case where the potential
$H$ appearing in the dynamical system for $m(z)$ is not a polynomial
in $m$. While not directly relevant for the hydrodynamic equations
studied in this paper for the Vicsek and active Ising model, such
Hamiltonians cannot be ruled out and may arise, for instance, from a
density-dependence of the coefficient $a_4$ in the phenomenological
hydrodynamic equations~\eqref{eq:hydroCC_m}.

\subsection{Vicsek like equations}
\label{sec:hydrovic}
Let us consider the hydrodynamic
equations~(\ref{eq:bertin-hydro-rho}-\ref{eq:bertin-hydro-m})
introduced by Bertin and co-workers to describe a simplified version
of the Vicsek model~\cite{Bertin}. These equations have the same
structure as the phenomenological equations
(\ref{eq:hydroCC_rho}-\ref{eq:hydroCC_m}) with two additional gradient
terms $\frac{\kappa}{2} \nabla(|\vec m|^2)$ and $-\kappa (\nabla\cdot
\vec m)\vec m$.  Importantly, all the coefficients $\gamma$, $\nu$,
$\kappa$, $\mu$ and $\zeta$ depend on the density.

We follow the same approach as before. Looking for propagative solutions
invariant in the transverse direction we set $m_y=0$, write $m_x=m$,
and go to the comoving frame $z=x-ct$, to obtain
\begin{align}
  \label{eq:bertin-ode-rho}
  c&\dot \rho-v_0 \dot m=0\\
  \label{eq:bertin-ode-m}
  \nu \ddot m&+\left(c-\frac{v_0^2}{2 c}\right)\dot m-\gamma m\dot m-\frac{v_0}{2}\dot \rho+\mu m-\zeta m^3=0
\end{align}
Note that after setting $m_y=0$ the two $\kappa$ gradient terms of
Eq.~(\ref{eq:bertin-hydro-m}) cancel each other.

As before, Eq.~(\ref{eq:bertin-ode-rho}) directly yields
\begin{equation}
  \rho(z)=\rho_g+\frac{v_0}{c}m(z)
\end{equation}
As we show in Appendix~\ref{sec:Vicsekode},
Eq.~(\ref{eq:bertin-ode-m}) can also be greatly simplified using the
explicit density-dependence of its coefficients. Introducing
$\tilde\gamma=\gamma/\nu$, $\tilde\zeta=\zeta/\nu$, and writing
$\mu=\mu_1\rho-\mu_2$ and $\nu^{-1}=\nu_1\rho+\nu_2$, one obtains the
second-order ordinary differential equation
\begin{equation}
  \label{eq:bertin-ode}
  \ddot m+(\alpha-\xi m)\dot m-a_2 m+ a_3 m^2 -a_4  m^3=0
\end{equation}
where the coefficients are all function of $c$ and $\rho_g$
\begin{align}
  \label{eq:bertin-coeffs}
  \alpha&=\left(c-\frac{v_0^2}{2c}\right)\left(\nu_1\rho_g+\nu_2\right) \qquad \xi=\frac{v_0^3}{2c^2}\nu_1+\tilde\gamma \\
  a_2&=\mu_2\nu_2+(\mu_2\nu_1-\mu_1\nu_2)\rho_g-\mu_1\nu_1\rho_g^2 \nonumber\\
  a_3&=\frac{v_0}{c}\left(2\rho_g\mu_1\nu_1+\mu_1\nu_2-\mu_2\nu_1\right)\nonumber\\
  a_4&=\tilde\zeta-\frac{v_0^2}{c^2}\mu_1\nu_1\nonumber
\end{align}
Interestingly, Eq.~(\ref{eq:bertin-ode}) has exactly the same form as
Eq.~\eqref{eq:ode}---the dynamical system stemming from the
phenomenological hydrodynamic equations studied in
Sec~\ref{sec:propag}---although with coefficients depending in a more
complicated way on $\rho_g$ and $c$. Their propagative solutions will
thus be the same, up to a slightly different organisation in the
$c,\rho_g$ plane.

However, an important difference between the Vicsek and phenomenological
hydrodynamic equations, is the scaling of the magnetization
with density in the ordered phase. The Vicsek hydrodynamic
equations~(\ref{eq:bertin-hydro-rho}-\ref{eq:bertin-hydro-m}) of
Bertin and co-workers indeed predict that the homogeneous ordered
solution satisfies
\begin{equation}
  \label{eq:m-to-zero_vicsek}
  \frac{|\vec m_0|}{\rho_{0}}=\frac{1}{\rho_{0}}\sqrt{\frac{\mu}{\zeta}}\xrightarrow[\rho_{0}\to\infty]{} \sqrt{\frac{\mu_1\nu_1}{\tilde \zeta}}
\end{equation}
which is consistent with what is observed in microscopic models such
as the Vicsek model. On the contrary, because the coefficient $a_4$ in
Eq.~\eqref{eq:hydroCC_m} does not depend on density, the
phenomenological equations~(\ref{eq:hydroCC_rho}-\ref{eq:hydroCC_m})
would yield $\frac{m_{0}}{\rho_{0}} \to 0$ as $\rho_{0}$ increases. In
this region of parameter space, which is not the main focus of this
paper, the Vicsek hydrodynamic equations are thus more faithful to the
phenomenology of microscopic models studied in Sec.~\ref{sec:micro}
than the phenomenological
equations~(\ref{eq:hydroCC_rho}-\ref{eq:hydroCC_m}). As we show in
Appendix~\ref{sec:a4ofrho}, however, these phenomenological equations
can recover a scaling akin to that of the Vicsek model if the
coefficient $a_4$ of Eq.~(\ref{eq:hydroCC_m}) is allowed to depend on
density. This leads to a slightly more complicated dynamical
system~\cite{Caussin} that we study in the appendix for completeness.

\begin{figure}
  \includegraphics[width=1\columnwidth]{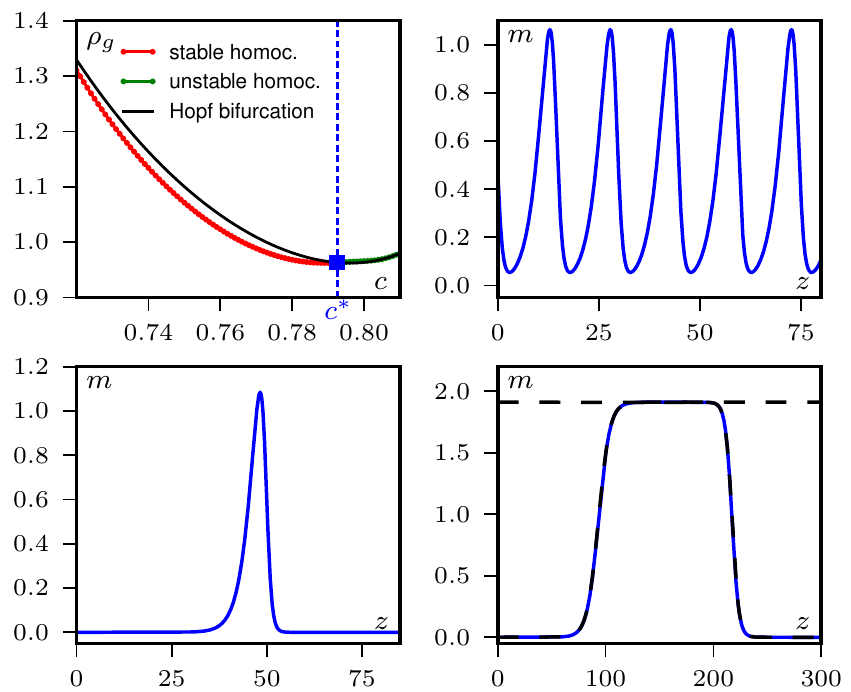}
  \caption{{\bf Top left}: Phase diagram for the propagative solutions of the
    hydrodynamic equations of Bertin \etal
    Eq.~(\ref{eq:bertin-ode}). The color code and different phases are
    the same as for the $a_4=\text{cst}$ case
    Fig.~\ref{fig:rhog-c}. We obtain the same three types of
    inhomogeneous profiles: periodic, homoclinic and heteroclinic.
    $\sigma=0.7$.}
  \label{fig:rhog-c_vicsek}
\end{figure}

Coming back to the dynamical system~\eqref{eq:bertin-ode}, following
the same method as that introduced in Sec.~\ref{sec:propag}, one can
derive analytical expressions for the Hopf bifurcation line
$\rho_g^H(c)$ and the speed $c^*$ where the bifurcation becomes
subcritical. We show in Fig.~\ref{fig:rhog-c_vicsek} the phase diagram
in the ($c$, $\rho_g$)-plane and examples for the three types of
inhomogeneous trajectories. Again an exact solution for the
heteroclinic trajectory can be derived and (the dashed lines in
Fig.~\ref{fig:rhog-c_vicsek}) is found at speed $c^h=c^*$. As
expected, there is no qualitative difference with the simpler case
studied in Sec.~\ref{sec:existence}.

\subsection{Active Ising model equations}
\label{sec:hydroising}
The active-Ising hydrodynamic equations~\cite{AIM} seem {\it a priori}
different since they contain none of the non-linear gradient terms
found in the phenomenological and Vicsek hydrodynamic
equations. However, these terms are in fact generated by the diffusion term in
the dynamical equation~(\ref{eq:ising-hydro-rho}) for the density
field~\cite{AIM}, so that we will once again recover the same types of
propagative solutions.

Looking for  stationary profiles in the comoving frame $z=x-ct$,
Eq.~(\ref{eq:ising-hydro-rho}) and~(\ref{eq:ising-hydro-m}) reduce to
\begin{align}
  \label{eq:ising-ode-rho} D&\ddot \rho+c\dot \rho-v_0 \dot m=0 \\
  \label{eq:ising-ode-m} D& \ddot m+c\dot m-v_0 \dot \rho+2\left(\beta{}-1-\frac{r}{\rho}\right)m -\alpha\frac{m^3}{\rho^2}=0
\end{align}
Equation~\eqref{eq:ising-ode-rho} can be solved by expanding $\rho$ in
gradients of $m$. Introducing the ansatz
\begin{equation}
  \label{eq:ising-gradient-exp}
  \rho(z)=\rho_g+\sum_{k=0}^\infty \alpha_k \frac{d^k m}{dz^k}
\end{equation}
into Eq.~(\ref{eq:ising-ode-rho}), and solving order by order, we get
the following recursion relation
\begin{equation}
  \label{eq:ising-recursion}
  a_0=\frac{v_0}{c} \qquad a_{k+1}=-\frac{D}{c}a_k
\end{equation}
from which we obtain
\begin{equation}
  \label{eq:ising-solution-rho}
  \rho(z)=\rho_g+\frac{v_0}{c}\sum_{k=0}^\infty \left(-\frac{D}{c}\right)^k \frac{d^k m}{dz^k}
\end{equation}
In the following we retain only the first order terms in gradient:
\begin{equation}
  \label{eq:ising-gradient-1storder}
  \rho(z)=\rho_g+\frac{v_0}{c}m(z)-\frac{D v_0}{c^2}\dot m(z)+\frac{D^2v_0}{c^3}\ddot m(z)
\end{equation}

To simplify Eq.~(\ref{eq:ising-ode-m}) we linearize around the density
$\varphi_g=r/(\beta{}-1)$ where the disordered profile becomes linearly
unstable. As shown in~\cite{AIM}, this is a good approximation when
$\beta \gtrsim 1$ because $\varphi_g\to\infty$ while $\rho-\varphi_g$
remains finite for inhomogeneous solutions. We then obtain
\begin{equation}
  \label{eq:ising-ode-m-linear}
  D \ddot m+c\dot m-v_0 \dot \rho+\frac{2r}{\varphi_g^2}(\rho-\varphi_g)m -\alpha\frac{m^3}{\varphi_g^2}=0
\end{equation}

Finally, inserting Eq.~(\ref{eq:ising-gradient-1storder}) in the
previous equation we obtain a second-order ordinary differential
equation with exactly the same terms as those obtained from the
phenomenological and Vicsek hydrodynamic equations, Eq.~(\ref{eq:ode})
and (\ref{eq:bertin-ode}):
\begin{equation}
  \label{eq:ising-ode-m-linear_bis}
  \tilde D \ddot m+\left(c-\frac{v_0^2}{c}-\xi m\right)\dot m-a_2 m+a_3m^2-a_4 m^3=0
\end{equation}
with
\begin{align}
  \label{eq:ising-coeffs}
  &\tilde D=D\left(1+\frac{v_0^2}{c^2}\right) \qquad \xi=\frac{4rDv_0}{(c^2+v_0^2)\varphi_g^2}\\
  &a_2=\frac{2r(\varphi_g-\rho_g)}{\varphi_g^2}\qquad a_3=\frac{2 r v_0}{c\rho_g^2} \qquad a_4=\frac{\alpha}{\rho_g^2}
\end{align}

This equation has again the same qualitative behavior as in the
phenomenological theory and the same three types of inhomogeneous
solutions are found with the same organization in the ($c$, $\rho_g$)
parameter space.

\section{Linear stability of the propagative solutions in the 1D  hydrodynamic equations}
\label{sec:pdestability}
In sections~\ref{sec:propag} and~\ref{sec:CGequations}, we have
found and classified all the propagative solutions of different
hydrodynamic equations. We have shown that in all cases three types
of such solutions {\it exist}: periodic patterns of finite-size
bands, solitary band solutions, and phase-separated solutions.
These solutions were found as stable limit cycles, homoclinic, and
heteroclinic orbits $m(z)$ of the reduced dynamical system
\eqref{eq:ode}.  This study does {\it not} tell us anything about
the local and a fortiori global {\it stability} of these solutions
as solutions $m(x-ct)$ and $\rho(x-ct)$ of the original hydrodynamic
partial differential equations. Indeed, a stable orbit of the
reduced dynamical system can very well be unstable to spatiotemporal
perturbations when re-expressed as an inhomogeneous propagative
solution of the hydrodynamic equations.  For example, consider the
fixed point at $m=m_-$ which is stable for $\rho_g>\rho_g^H$, in the
region I and III in Fig.~\ref{fig:rhog-c}. Without performing any
calculation, we know that the corresponding homogeneous solution
($\rho_0=\rho_g+\frac{v_0}{c}m_-$, $m_0=m_-$) is unstable in the
hydrodynamic equations because it lies inside the spinodal region
where no homogeneous solution is linearly stable.

In this section we study the linear (local) stability of these
solutions at the hydrodynamic level.  Although, it is a well-defined
linear problem, determining the stability of inhomogeneous solutions
of nonlinear partial differential equations cannot be done
analytically even in the (rare) cases where these solutions are known
analytically. A direct numerical study is possible (for an example,
see, e.g. \cite{NBSTAB}), but is rather tedious, all the more so as we
deal with 2D hydrodynamic equations.  In the following, we study
mostly the linear stability of the hydrodynamic equations reduced to
one dimension (that of propagation), using a simple numerical
procedure explained below.  For an account of some fully 2D
preliminary investigation, see the end of this section.

\subsection{Numerical procedure}

We investigated numerically the
stability of the propagative solutions of the phenomenological
hydrodynamic equations Eq.~(\ref{eq:hydroCC_rho}-\ref{eq:hydroCC_m}) 
reduced to one space dimension, that of propagation. To do so, we select a
solution of the corresponding classical mechanics
problem.~(\ref{eq:ode}) and use it as initial condition of the
numerical integration of Eqs.~(\ref{eq:hydroCC_rho}) and
(\ref{eq:hydroCC_m}).

The numerical integration is done using a semi-spectral algorithm, the
linear terms being computed in Fourier space, the non-linear ones in
real space and a semi-implicit time-stepping. This method, where the
fields $\rho$ and $m$ are represented by their $N$ first Fourier modes
(with $N$ large enough that the simulation has converged), is well
suited to simulate systems with diffusion terms and periodic boundary
conditions.

Preparing the initial condition of the hydrodynamic equations always
brings in discretization errors. Indeed, because of the periodic
boundary conditions in the hydrodynamic equations, we need to select a portion of the
solution $m(z)$ which is a multiple of the period. This is done with
an error of order $dx$, the space discretization step used in the
numerical integration of~(\ref{eq:hydroCC_rho}) and
(\ref{eq:hydroCC_m}).

\begin{figure}
  \includegraphics[width=1\columnwidth]{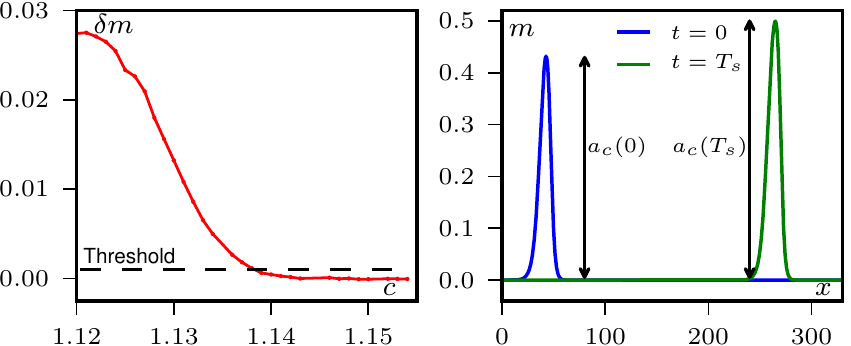}
  \caption{Stability of homoclinic orbits in the hydrodynamic equations. $\Delta a_c$ is
    the difference in the amplitude of the solution measured between
    $t=0$ and $t=T_s$. The solution is declared to be linearly stable
    if $\delta m(T_s)$, defined in Eq.~\eqref{eq:pdeltam} is smaller than a threshold that we take at
    $10^{-3}$.  $v_0=\lambda=\xi=a_4=\varphi_g=1$, $c=1.125$ (right)}
  \label{fig:stability_homoclines}
\end{figure}

Accordingly, we observe a rapid relaxation at short times due to the
discretization errors. Subsequently, we find that the original
solution is either stable at long times or is quickly destabilized and
converges to another solution. To analyze systematically the stability
of the propagative solutions, we defined a quantitative criterion for
the stability: We choose a time $T_s=2000$ much larger than the
relaxation time of the initial perturbation (which happens in a time
$\sim 100$) but not too large to test solely the linear stability
regardless of a possible long-time coarsening dynamics that could be
induced by numerical noise. We then measure the amplitude of the
solution $|\Delta m|(t)$ defined in Eq.~(\ref{eq:deltam}) as a
function of time. If
\begin{equation}\label{eq:pdeltam}
\delta m(T_s)\equiv |\Delta m(T_s)-\Delta m(0)|
\end{equation}
is smaller than $10^{-3}$, the solution is said to be stable and
unstable otherwise (see Fig.~\ref{fig:stability_homoclines}).
This protocol does not give exact answers to the question of
linear stability, since in particular the small but finite initial
perturbations may take the initial condition out of the basin of
attraction of a (stable) solution. But the results presented below
are relatively robust to changing our numerical resolution and the
conditions used to decide stability, and we are thus confident that
they represent well the 'true' subset of linear stable solutions.

The precise criterion does not matter much for the results since we
find an abrupt transition from stable to unstable trajectories (see
Fig.~\ref{fig:stability_homoclines}) which is visible on a variety of
observables (norm of the fields, period, max and min values, etc).

\subsection{Results}
\begin{figure}
  \includegraphics[width=0.8\columnwidth]{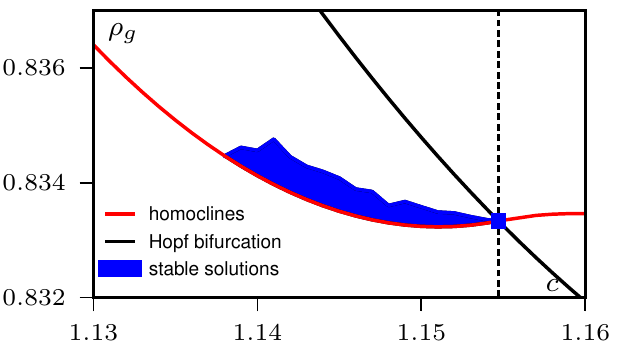}
  \caption{The blue region is the subset of admissible propagative
    solutions which are linearly stable. None of the solutions for
    $c>c^*$ is stable in the hydrodynamic
    equations. $v_0=\lambda=\xi=a_4=\varphi_g=1$. We use system sizes
    $L_x=300$ or more (adapted to fit the solution) 
    with resolution $dx=0.5$ and $dt=0.1$.}
  \label{fig:stabilityPDE}
\end{figure}

\begin{figure}
  \includegraphics[width=1\columnwidth]{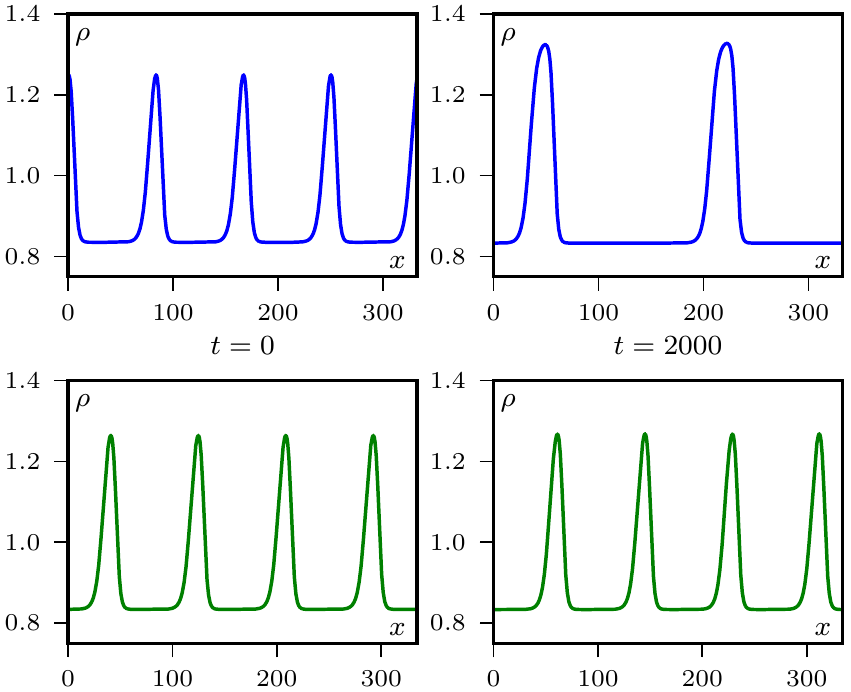}
  \caption{Evolution of two propagative solutions in the hydrodynamic
    equations ~(\ref{eq:hydroCC_rho}-\ref{eq:hydroCC_m}). The initial
    conditions are stable limit cycles of the dynamical system
    \eqref{eq:ode} for $m(z)$. 
{\bf Top row}:
    unstable solution with $c=1.135$, $\rho_g=0.8351$. {\bf Bottom
      row}: stable solutions with $c=1.14$, $\rho_g=0.8341$. System
    size $340\times 100$, $dt=0.1$,
    $dx=0.4$. $v_0=\lambda=\xi=a_4=\varphi_g=1$.}
  \label{fig:profilesPDE}
\end{figure}

Figure~\ref{fig:stabilityPDE} contains one of the central results of
our paper: only a very small subset of the propagative solutions are
stable at the hydrodynamic level. However this subset includes the
three possible types of trajectories: periodic bands, solitonic bands,
and phase-separated profiles. The linearly stable solutions are all
found in the region of the ($c$, $\rho_g$) plane close to the
heteroclinic trajectory and close to the line of homoclinic
orbits. Examples of stable and unstable solutions are shown in
Fig.~\ref{fig:profilesPDE}.

To understand why only the vicinity of the heteroclinic and homoclinic
solutions is stable, one can argue that the solutions must have a
large enough amplitude to be dynamically stable. Indeed, the periodic
solutions oscillate around $m=m_-$ which lies inside the spinodal
region of the hydrodynamic equations, where no homogeneous solutions is
stable. Small-amplitude oscillations around this point should thus
also by dynamically unstable and only large enough amplitude cycles,
found near the homoclinic line and the heteroclinic trajectories, are
stable.

Note that the region where stable solutions are found has a rather
rough boundary in Fig.~\ref{fig:stabilityPDE}. This is most probably
an artefact due to the initial discretization error which is not
controlled and varies from one propagative solution to another. Close
to the threshold of linear instability, this can easily make a
(linearly) stable trajectory unstable.

We report finally on fully 2D simulations performed on
rectangular boxes of width $L_y=100$, with small noise added to each
grid point on a 1D solution extended trivially along $y$. These
yielded essentially no change with respect to the results presented
above. While this encourages us to believe that no unstable mode has
components along $y$, and thus that the subset determined above
corresponds to the linearly stable solutions of the full 2D
hydrodynamic equations, we remain cautious. As a matter of fact,
recent results obtained in the case of hydrodynamic equations for
active nematics have revealed the existence of (very) long
wavelength instability along the homogeneous direction of band
solutions.  \cite{NEMATIC} A similar investigation is left for
future studies.

\section{Coarsening in the hydrodynamic equations}
\label{sec:coarsening}
The two-dimensional subset of propagative solutions that are
linearly stable still contains an infinity of solutions, including
smectic patterns, solitary bands, and phase-separated profiles. We
can thus wonder whether a single solution is selected in the
hydrodynamic equation starting from a random initial condition.

One can obtain some insight into this question by studying the lines of
propagative solutions having a given average density $\rho_0$, shown
in Fig.~\ref{fig:fixed-rho0}. Indeed, $\rho_0$ is conserved in the
hydrodynamic equations and any stable propagative solutions has to lie
on such a line. As discussed earlier, the larger the amplitude of a
solution, the faster the propagation. We can thus expect that if
several travelling bands coexist, they will encounter and merge
because of their different speeds. This will in turn increase the
sizes of the surviving objects and hence their speed. This mechanism
would naturally lead the system toward the heteroclinic solution, {\it
  i.e.}, to the phase-separated state.

\begin{figure*}
  \includegraphics[width=1\textwidth]{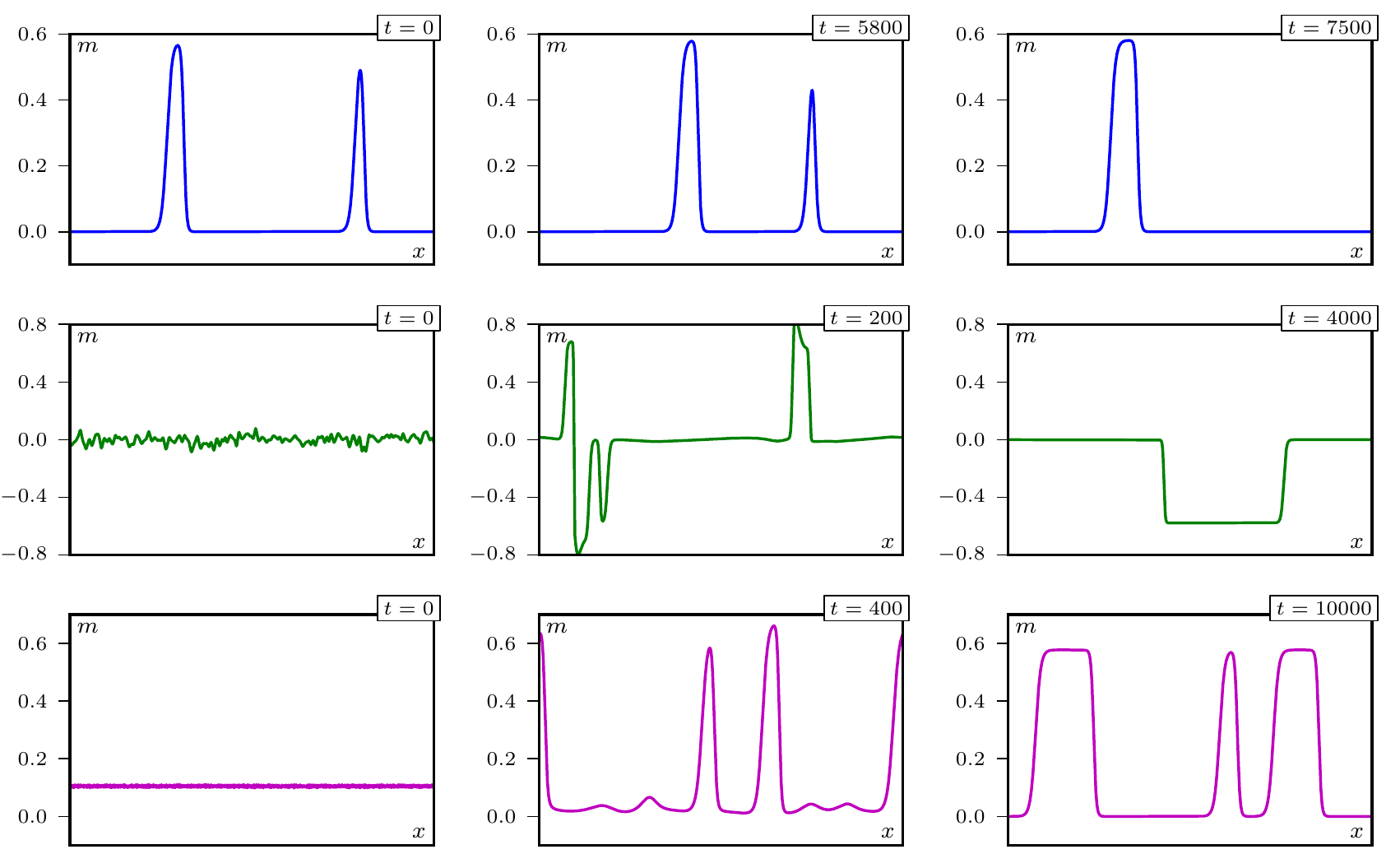}
  \caption{Coarsening in the phenomenological hydrodynamic
    equations~(\ref{eq:hydroCC_rho}-\ref{eq:hydroCC_m}) from
    three different initial conditions. {\bf Top}: Two homoclinic
    trajectories glued together. {\bf Middle}: Disordered initial
    condition. {\bf Bottom}: Ordered initial condition.  System size
    $500\times 100$ (top and bottom), $1000\times 100$ (middle). The
    profiles are averaged along the $y$-direction in which the system
    is invariant.  $v_0=\lambda=\xi=a_4=\varphi_g=1$, $dt=0.1$,
    $dx=0.1$.}
  \label{fig:coarsening}
\end{figure*}

We checked this scenario by simulating the phenomenological
hydrodynamic equations~(\ref{eq:hydroCC_rho}-\ref{eq:hydroCC_m}) with
three different initial conditions, as shown in
Fig.~\ref{fig:coarsening}:
\begin{itemize}
\item We build an initial condition made of two homoclines glued
  together, both solutions of the differential equation~\eqref{eq:ode}
  for different speed $c$ and inside the stability domain of
  Fig.~\ref{fig:stabilityPDE}. To avoid discontinuities we interpolate
  smoothly between the gas densities of the two solutions using a
  hyperbolic tangent function.  We then observe that the two
  travelling bands get closer until, when close enough (though not in
  contact),  the smaller one evaporates, its mass
  being transfered to the second band. The final solution is
  hence a single larger isolated band.

\item Starting from a random ordered solution with constant $\rho$ and
  a magnetization $m$ fluctuating around $m_0\neq 0$, several bands
  form at short times that all go in the same direction. The system
  then coarsens because of the speed differences between the liquid
  droplets until only ``phase-separated domains'', that all have the
  same speed, remain. In practice, because the band speeds can be very
  close, a final state with a single phase-separated profile may not
  be reached within the time-scales of our simulations and a precise
  study of this coarsening regime is beyond our numerical capacities.
\item Starting from a disordered initial condition with $m(x)$
  fluctuating around $0$ and a density inside the spinodal region,
  many domains of positive and negative magnetization form. These
  objects then encounter and merge yielding a rapid coalescence
  process. Because of the periodic boundary conditions, this process
  typically yields a single phase-separated state for the system
  sizes we considered. For larger systems, the coalescence process
  could result in  a larger number of bands propagating in the same
  direction. We should then observe the same type of coarsening as in
  the second case discussed above.
\end{itemize}
At the level of hydrodynamic equations, the fact that larger
ordered domains travel faster leads to a natural coarsening towards
the phase-separated states. This coarsening relies both on a coalescence
process and, when bands travelling in the same direction are close
enough, on a ripening during which a smaller band evaporates and is
``swallowed'' by its larger neighbour.

Note that we studied the stability of propagative solutions and their
coarsening process using the phenomenological hydrodynamic
equations~(\ref{eq:hydroCC_rho}-\ref{eq:hydroCC_m}). While the precise
results will depend on the set of hydrodynamic equations under study,
our simulations of the Vicsek and Ising hydrodynamic
equations,~(\ref{eq:bertin-hydro-rho}-\ref{eq:bertin-hydro-m}) and
(\ref{eq:ising-hydro-rho}-\ref{eq:ising-hydro-m}), did not suggest any
qualitative difference. Their comprehensive investigation, which is
beyond the scope of this paper, would nevertheless be interesting,
especially to quantifiy the role played by the non-linearities in the
vectorial hydrodynamic equations~\eqref{eq:bertin-hydro-m}.

\section{Phase diagram of coarse-grained hydrodynamic equations}
\label{sec:phase-diagrams-hydro}

\begin{figure*}
  \includegraphics[width=0.4\textwidth]{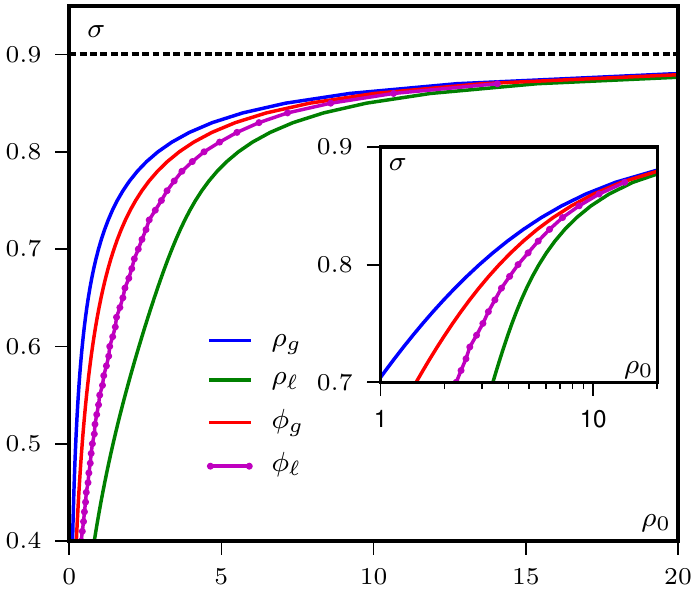}
  \includegraphics[width=0.4\textwidth]{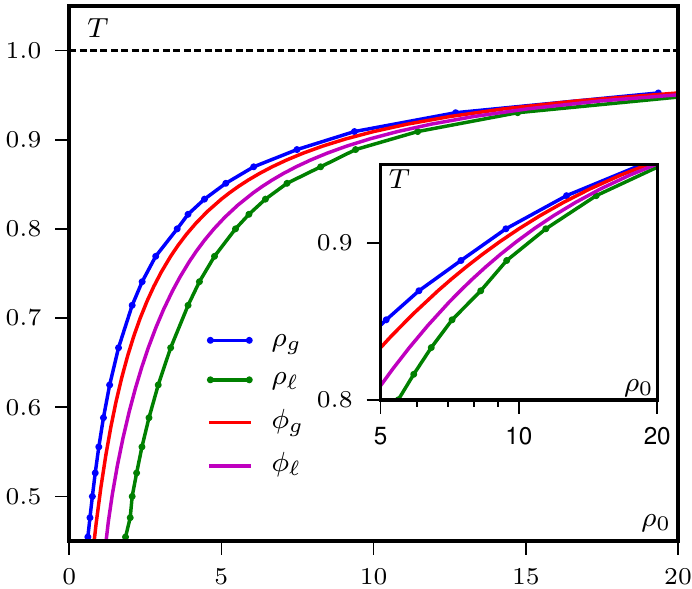}
  \caption{Phase diagram of the Visek (left) and Ising (right)
    hydrodynamic equations. The spinodal lines $\varphi_g$ and
    $\varphi_\ell$ are known exactly except $\varphi_\ell$ in the
    Vicsek hydrodynamic equation which we computed numerically by
    looking at the stability of systems of size $50\times 50$. The
    binodals $\rho_g$ and $\rho_\ell$ are the coexisting densities of
    phase-separated (heteroclinic) solutions as explained in the text.
    The dashed line indicate the asymptote above which only disordered
    solutions exist. Insets are close-ups on the high-density regions
    with a logarithmic scale on the $x$-axis. Parameters:
    $v_0=D=1=r=1$ (Ising), $v_0=1$ (Vicsek).}
  \label{fig:phase-diagram_PDE}
\end{figure*}

We can now construct the full phase diagram for the Ising and Vicsek
hydrodynamic equations in the temperature-density ensemble (or
noise-density for Vicsek), Fig.~\ref{fig:phase-diagram_PDE}. It is
composed of two spinodal lines, which are the limit of linear
stability of the homogeneous disordered and ordered profiles, and two
binodal lines outside which inhomogeneous solutions cannot be
observed.

\subsection{Spinodal lines}
The spinodal lines can be determined analytically from a linear
stability analysis, as was previously done for the Vicsek
model~\cite{Bertin} and the active Ising model~\cite{AIM}. The lower
spinodal line $\varphi_g$ is simply the density at which the
coefficient of the term linear in $m$ in the hydrodynamic equations
changes sign (in our phenomenological equation, when $a_2=0$). It
reads for the Ising and Vicsek hydrodynamic equations
\begin{align}
  \label{eq:phig} \varphi_g^I(T)&=r/(\beta-1) \\
  \varphi_g^V(\sigma)&=\frac{\mu_2}{\mu_1}=\frac{\pi(1-e^{-\sigma^2/2})}{4(e^{-\sigma^2/2}-2/3)}
\end{align}
where $\beta=1/T$. The higher spinodal $\varphi_\ell$ can be determined
exactly for the Ising hydrodynamic equations
\begin{equation}
  \label{eq:phil} \varphi_\ell^I(T)=\varphi_g\frac{v_0\sqrt{\alpha\left(v_0^2\kappa+8 D(\beta-1)^2\right)}+v_0^2\kappa+8 D \alpha (\beta-1)}
  {2v_0^2\kappa +8  D \alpha (\beta-1)}
\end{equation}
where $\kappa=2+\alpha-2\beta$ and $\alpha=\beta^2(1-\beta/3)$. For
the Vicsek hydrodynamic equation, the exact determination of
$\varphi_\ell$ is much more cumbersome~\cite{Bertin}. In
Fig.~\ref{fig:phase-diagram_PDE} we show the line
$\varphi_\ell^V(\sigma)$ computed numerically by simulating the Vicsek
hydrodynamic equations at different densities in the homogeneous
ordered state and looking at the growth of a small perturbation.

\subsection{Binodal lines}
The binodal lines $\rho_g$ and $\rho_\ell$ are defined as the minimum
and maximum global densities beyond which inhomogeneous propagative profiles
cannot be observed in simulations of the hydrodynamic equations. As
explained in Sec.~\ref{sec:fixed-rho0}, at the heteroclinic trajectory
the size of the liquid and gas domains are arbitrary so that
phase-separated solutions can have any density in the range
$[\rho_g^h,\rho_\ell^h]$. We find that, for all parameters we tested,
$\rho_\ell^h=\rho_\ell$, i.e., no other stable solution has a larger
density than the liquid domain of the heteroclinic solution.

\begin{figure}
  \includegraphics[width=1\columnwidth]{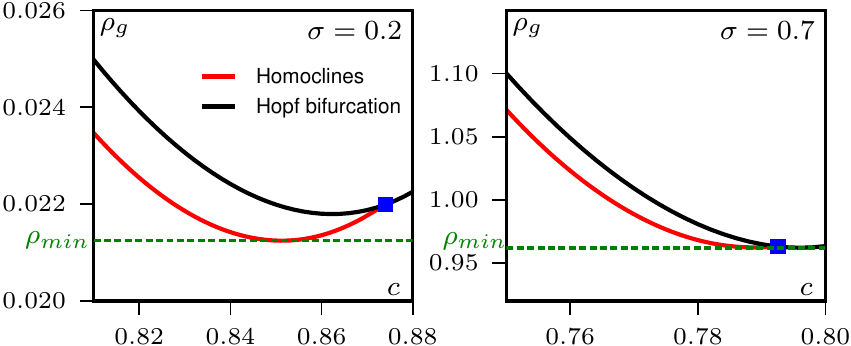}
  \caption{Propagative solutions of the Vicsek hydrodynamic equations
    for $\sigma=0.2$ and $0.7$. The solution with the lower mean
    density $\rho_{\text{min}}$ is found at the minimum of the line of
    homoclinic solutions. At low noise, it does not coincide with the
    heteroclinic solution (blue squares). $v_0=1$.}
  \label{fig:minimum_homocline}
\end{figure}

The situation is more subtle for the lower binodal $\rho_g$. Depending
on the external parameters, the line of homoclinic trajectories
solution of the dynamical system is not always monotonous as a
function of $c$. For example, in the Vicsek hydrodynamic equation it
is not monotonous at low noise, as shown in
Fig.~\ref{fig:minimum_homocline}. In this case the minimum density
accessible to propagative solutions is the minimum of the homoclinic
line. This solution need not be stable in the hydrodynamic equations
so that one should repeat the stability analysis done in
Sec.~\ref{sec:pdestability} for each value of the noise to determine
exactly the binodal line. For simplicity, the lower binodal of the
Vicsek hydrodynamic equation shown in Fig.~\ref{fig:phase-diagram_PDE}
is the gas density read from the phase-separated profile, which is
true at high noise and a good approximation of $\rho_g$ at lower
noise values.

For the Vicsek hydrodynamic equations, the coexisting densities of
the heteroclinic trajectory are known exactly whereas in the Ising
case they can be determined analytically only after linearizing
Eq.~(\ref{eq:ising-ode-m}) around $\rho=\varphi_g$. The binodal lines
in the phase diagram of the Ising hydrodynamic equations shown in
Fig.~\ref{fig:phase-diagram_PDE} are thus determined by integrating
numerically the hydrodynamic equations and measuring the density of
the liquid and gas domain of a phase-separated solution whereas we
plot the analytical solution in the Vicsek case.

\section{Conclusion}

Before summarizing our results, let us discuss how the study of the
hydrodynamic equations presented in this article compares with the
phenomenology of the microscopic models. The phase diagrams of the
Ising and Vicsek hydrodynamic equations shown in
Fig.~\ref{fig:phase-diagram_PDE} are qualitatively similar. They are
also consistent with the phase diagrams of the microscopic models
shown in Fig.~\ref{fig:diagrams_micro}. The hydrodynamic equations
thus provide a picture which is consistent with the liquid-gas
framework discussed in Sec.~\ref{sec:LGT}. For instance, the asymptote
at finite noise (or temperature) can again be seen as the simplest way
of forcing the system to cross a transition line to go from its gas to
liquid phase.

The comparison between the phase-separated regions of the microscopic
models and hydrodynamic equations is however more subtle. We have
indeed shown, using the phenomenological hydrodynamic
equation~(\ref{eq:hydroCC_rho}-\ref{eq:hydroCC_m}), that the
coarsening leads, at the hydrodynamic level, to the phase-separated
solution. This is true in the active Ising model but not in the Vicsek
model, which only shows micro-phase separation and thus a coarsening
leading to a periodic solution (see Fig.~\ref{fig:bands_micro}). This
difference between microscopic models and hydrodynamic equations is
however not surprising since it was recently shown that fluctuations
are essential to account for the selection of micro-phase separated
profiles in the Vicsek model~\cite{Microphase}.

More precisely, phase-separated and micro-phase-separated
solutions are both linearly stable in the hydrodynamic equations for
vectorial order parameters. As noise is added to these equations,
though, large bands are destabilized and break in periodic array of
finite-size bands, in agreement with what is seen in microscopic
simuations of the Vicsek model~\cite{Microphase}. We can now
tentatively put together these results to propose a coarsening process
that would lead to the micro-phase separated states seen in the Vicsek
model. Starting from a profile with many coexisting bands, the
coarsening \textit{at fixed $\rho_0$} would tend to lead to the
formation of larger and larger bands. This coarsening would however be
arrested by the fact that the fluctuations in the Vicsek model set a
maximal size beyond which bands are (non-linearly) unstable. How this
size is selected however remains to be determined.

All in all, we have shown in this paper that the hydrodynamic
equations describing polar flocking models generically share the same
propagative solutions. We found three types of solutions that are all
observed in microscopic models: periodic orbits that describe
microphase separation, homoclinic orbits describing solitonic objects
and heteroclinic trajectories that correspond to phase
separation. Only a subset is stable in the hydrodynamic equations, but
it still contains the three types of solutions.

The coarsening in the hydrodynamic equations favors the fastest solutions which are also
the largest patterns. It thus leads naturally to the phase-separated
solution which controls the phase diagram of the hydrodynamic equations. The same
behavior is observed in the microscopic active Ising model whereas one
can understand the microphase separation of the Vicsek model as an
arrested coarsening, the largest pattern being non-linearly unstable
to the (giant) fluctuations of the liquid phase.

\begin{acknowledgments}
  We thank T. Dauxois, A. Peshkov, J. Toner, V. Vitelli for many
  illuminating discussions. This project was supported through ANR
  projects BACTTERNS and MiTra.
\end{acknowledgments}

\appendix{}

\section{Limit of small $D$}
\label{sec:smallD}

While we do not have an analytic solution for the homoclinic profiles,
lots of insight on their physics can be gained by studying the limit
of small diffusion coefficient $D$. This is most easily done by
introducing the auxiliary variable $Z=D \dot m+F(m)$ where
\begin{equation}
  \label{eq:F}
  F(m)=\left(c-\frac{\lambda v_0}{c}\right)m-\frac{\xi}{2}m^2
\end{equation}
such that $F'(m)=f(m)$. Our dynamical system can then be recast as
\begin{equation}
  \label{eq:ode_system_Z}
  \frac{d}{dz}\colvec{m}{Z}=\colvec{\frac{1}{D}(Z-F(m))}{-H'(m)}
\end{equation}

\begin{figure}
  \includegraphics[width=1\columnwidth]{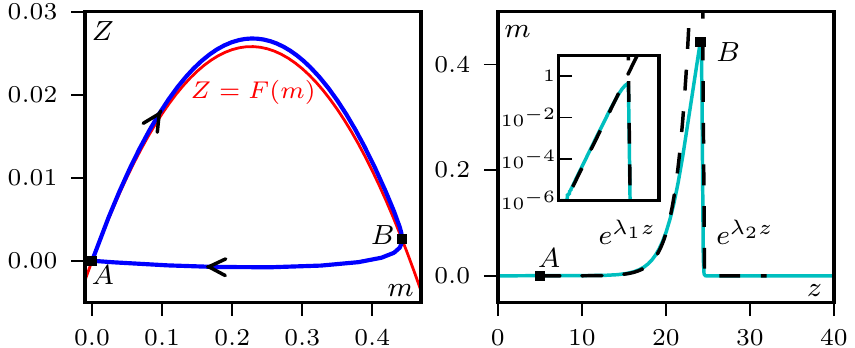}
  \caption{{\bf Left}: Phase portrait for the variables $m$ and
    $Z=D\dot m+F(m)$ where $F(m)$ is the red parabola. {\bf Right}:
    The corresponding trajectory. The dashed lines are the linear
    approximation around $m=0$. The inset in semi-log scale shows the
    exponential tail close to $m=0$. Parameters: $D=0.01$,
    $v_0=\lambda=\xi=a_4=\varphi_g=1$.}
  \label{fig:smallD}
\end{figure}

Let us consider a large-amplitude orbits that start close to $m=0$. An
example of such orbits and the corresponding phase portrait for the
($m$, $Z$) variables is shown in Fig.~\ref{fig:smallD}. When $D$ is
small, $\dot m$ relaxes quickly to zero so that $Z$ relaxes to the
parabola $Z=F(m)$ in a time $\sim D$.  Following the trajectory in
Fig.~\ref{fig:smallD}, starting from point A at $m\approx 0$, the
trajectory between A and B is above the parabola $F(m)$. The distance
with the parabola first increases when $m<m_-$, which implies
$H'(m)<0$ and $\dot Z>0$, and decreases afterwards when $m>m_-$. The
distance with the parabola stays of order $D$, set by the relaxation
time of $\dot m$. When $Z=f(m)$ at point $B$, $m$ relaxes to $m=0$ in
a time $\sim D$. This gives profiles with sharp leading fronts and
long exponential tails, which are indeed consistent with the profiles
seen at small temperature in the Vicsek model and its putative
hydrodynamic description~\cite{ChateGregoire,Bertin,Ihle}.

This picture is consistent with the fact that at leading order in $D$
the eigenvalues at point ($m=0$, $\dot m=0$) read
\begin{equation}
  \label{eq:eigenvalues_smallD}
  \lambda_1=\frac{\varphi_g-\rho_g}{c-\lambda v_0/c} \qquad \lambda_2=-\frac{c-\lambda v_0/c}{D}
\end{equation}
so that we have a slow unstable direction $\lambda_1$ and a fast
stable direction $\lambda_2$. These two eigenvalues indeed control a
large part of the trajectory, as shown in the right panel of
Fig.~\ref{fig:smallD}.

\section{Phenomenological equations with $a_4(\rho)$}
\label{sec:a4ofrho}
In the hydrodynamic equation~(\ref{eq:hydroCC_m}) the homogeneous ordered solutions
have a magnetization $|\vec m_0|=\sqrt{a_2/a_4}$. Thus, in the
simplified case studied in Sec.~\ref{sec:existence} where
$a_2=\rho-\varphi_g$ and $a_4=\text{cst}$, one observes that
\begin{equation}
  \label{eq:m-to-zero}
  \frac{|\vec m_0|}{\rho}=\sqrt{\frac{\rho-\varphi_g}{a_4 \rho^2}}\xrightarrow[\rho\to\infty]{} 0
\end{equation}

On the contrary we observe in the microscopic Vicsek and active Ising
models that, at large densities, $|\vec m_0|/\rho_0$ reaches a constant
$P_0\le 1$ (set by the noise intensity in the microscopic
models). This can be achieved by assuming that $a_4=(\rho P_0^2)^{-1}$
as in~\cite{Caussin} so that
\begin{equation}
  \label{eq:m-to-P0}
  \frac{|\vec m_0|}{\rho}=P_0\sqrt{\frac{\rho-\varphi_g}{\rho}}\xrightarrow[\rho\to\infty]{} P_0
\end{equation}

Repeating the same treatment as before, we look for 1d propagative
solutions with speed $c$ which are the solutions of
\begin{multline}
  \label{eq:ode_a4rho}
  D \ddot m+(c-\frac{\lambda v_0}{c}-\xi m)\dot m-(\varphi_g-\rho_g) m\\+\frac{v_0}{c}m^2 -\frac{m^3}{P_0^2(\rho_g+\frac{v_0}{c}m)}=0
\end{multline}
which can be written in the same form as Eq.~(\ref{eq:ode_potential}) with a
different potential
\begin{align}
  \label{eq:ode_potential_a4rho} D \ddot m&=-f(m)\dot m-H'(m) \\
  H(m)&=-\frac{c^3\rho_g^2}{P_0^2v_0^3}m+\left[\frac{c^2\rho_g}{2P_0^2v_0^2}-\frac{(\varphi_g-\rho_g)}{2}\right]m^2\nonumber\\
\label{eq:potential_a4rho} &-\left[\frac{c}{3P_0^2 v}-\frac{v_0}{3c} \right]m^3+\frac{c^4\rho_g^3}{P_0^2v_0^2}\log(v_0m+c\rho_g)\\
  \label{eq:friction_a4rho}f(m)&=c-\frac{\lambda v_0}{c}-\xi m
\end{align}

We find a behavior very similar to the case $a_4=\text{cst}$ discussed
in section~\ref{sec:propag}. Under the same constraints (C1-3), the
potential $H(m)$ has the same form with maxima in $m=0$ and $m=m_+$
and a minimum at $m=m_-$ given by
\begin{equation}
  \label{eq:a4-rho-mpm}
  m_\pm=\frac{cv_0P_0^2(2\rho_g-\varphi_g)\pm cP_0\sqrt{(P_0v_0\varphi_g)^2-4c^2\rho_g(\varphi_g-\rho_g)}}{2(c^2-P_0^2v_0^2)}
\end{equation}

\begin{figure}
  \includegraphics[width=1\columnwidth]{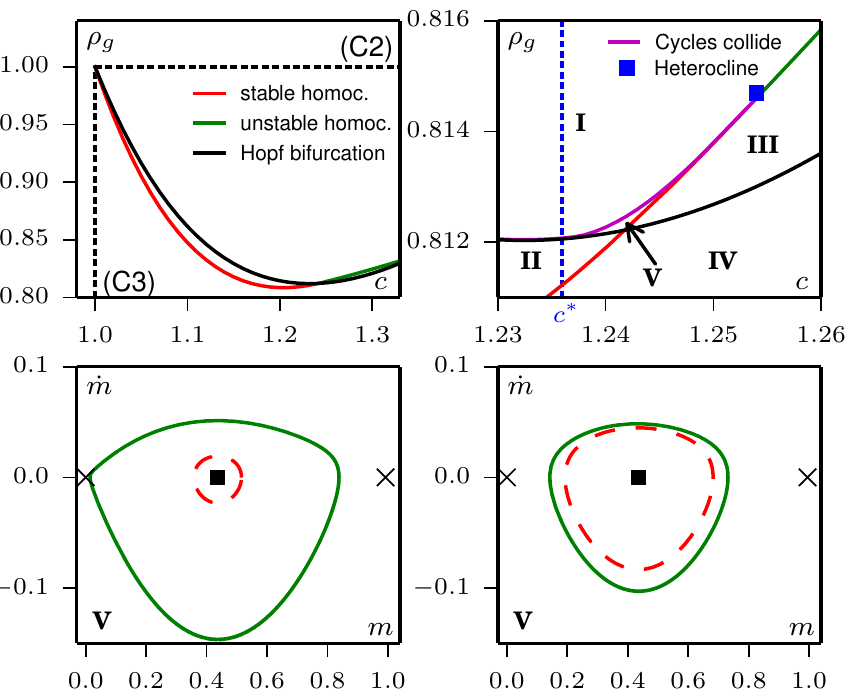}
  \caption{{\bf Top row}: Phase diagram for the propagative solutions
    of Eq.~(\ref{eq:ode_potential_a4rho}). The right plot is a zoom on
    the region $c^*<c<c^h$. Phase I to IV are the same as in the case
    $a_4=\text{cst}$ (see Fig.~\ref{fig:schema_hopf}). {\bf Bottom
      row}: Phase portraits of the solutions in phase V for the same
    $c=1.2421$ and $\rho_g=0.81228$ (left) and $\rho_g=0.81248$
    (right). When $\rho_g$ increases, the size of the stable cycle
    decreases while the size of the unstable cycle increases, until
    they collide as indicated on the phase diagram.
    $v_0=\lambda=\xi=P_0=\varphi_g=1$.}
  \label{fig:rhog-c_a4rho}
\end{figure}

One observes the same three types of trajectories already shown in
Fig.~\ref{fig:trajectories}: periodic, homoclinic and
heteroclinic. The phase diagram ($\rho_g$, $c$), shown in
Fig.~\ref{fig:rhog-c_a4rho}, is also similar to the case
$a_4=\text{cst}$ except in a very small region close to the
heteroclinic trajectory. We find again a 2-parameter family of
periodic orbits and a line of homoclinic solutions which terminates at
a unique heteroclinic trajectory.

As before, we can compute analytically the line $\rho_g^H(c)$ where
the Hopf bifurcation takes place and the speed $c^*$ such that the
bifurcation is supercritical for $c<c^*$ and subcritical for
$c>c^*$. The difference with the previous case is that the
heteroclinic trajectory, for which we do not have an exact solution
anymore, is not found at $c=c^*$ but at $c^h>c^*$. As a consequence,
we observe a new phase (shown as number V in
Fig.~\ref{fig:rhog-c_a4rho}) in which two limit cycles are found. A
large stable cycle surrounds a smaller unstable limit cycle which
itself encapsulates the stable fixed point $m=m_-$. As seen from the
phase portraits in Fig.~\ref{fig:rhog-c_a4rho}, when increasing
$\rho_g$ the size of the stable cycle decreases whereas the size of
the unstable one increases. Thus the upper limit of existence of this
phase (the magenta line in the phase diagram) is the moment where the
cycles collide and annihilate.

\section{Propagative solutions of the Vicsek hydrodynamic equations}
\label{sec:Vicsekode}
Here we show how the study of propagative solutions in the Vicsek
hydrodynamic equations leads to the same dynamical system as for the
phenomenological equations, but with coefficients depending in a
more complicated way on $\rho_g$ and $c$.

We start from the hydrodynamic
equations~(\ref{eq:bertin-hydro-rho}-\ref{eq:bertin-hydro-m}) in the
comoving frame $z=x-ct$:
\begin{align}
  \label{Beq:bertin-ode-rho}
  c&\dot \rho-v_0 \dot m=0\\
  \label{Beq:bertin-ode-m}
  \nu \ddot m&+\left(c-\frac{v_0^2}{2 c}\right)\dot m-\gamma m\dot m-\frac{v_0}{2}\dot \rho+\mu m-\zeta m^3=0
\end{align}
The coefficients in Eq.~(\ref{Beq:bertin-ode-m}) are greatly simplified
by dividing the equation by $\nu$, thus getting
\begin{equation}
  \label{Beq:bertin-ode-m-nu}
  \ddot m+\frac{1}{\nu}\left(c-\frac{v_0^2}{2 c}\right)\dot m-\tilde\gamma m\dot m-\frac{v_0}{2\nu}\dot \rho+\frac{\mu}{\nu} m-\tilde\zeta m^3=0
\end{equation}
where $\tilde\gamma=\gamma/\nu$ and $\tilde\zeta=\zeta/\nu$ depend
only on the noise $\sigma$ on the alignment interaction~\cite{Bertin}
\begin{align}
  \label{Beq:bertin-gammat}
  \tilde\gamma&=\frac{8}{\pi}\left(\frac{16}{15}+2e^{-2\sigma^2}-e^{-\sigma^2/2}\right) \\
  \label{Beq:bertin-zetat}
  \tilde\zeta&=\frac{64}{\pi^2}\left(e^{-\sigma^2/2}-\frac{2}{5}\right)\left(\frac{1}{3}+e^{-2\sigma^2}\right)
\end{align}
while $\mu$ and $\nu$ depend also on the density
\begin{align}
  \label{Beq:bertin-mu}
  \mu&=\frac{4 v_0\rho}{\pi}\left(e^{-\sigma^2/2}-\frac{2}{3}\right) -v_0\left(1-e^{-\sigma^2/2}\right)\\
  \label{Beq:bertin-nu}
  \nu^{-1}&=\frac{4}{v_0^2}\left[\frac{8\rho}{3\pi}\left(\frac{7}{5}+e^{-2\sigma^2}\right)+\left(1-e^{-2\sigma^2}\right)\right]
\end{align}

One can solve Eq.~(\ref{Beq:bertin-ode-rho}) to get
$\rho(z)=\rho_g+\frac{v_0}{c}m(z)$ as before and use it in
Eq.~(\ref{Beq:bertin-ode-m-nu}) to obtain the second-order ordinary
differential equation
\begin{equation}
  \label{Beq:bertin-ode}
  \ddot m+(\alpha-\xi m)\dot m-a_2 m+ a_3 m^2 -a_4  m^3=0
\end{equation}
where the coefficients are all function of $c$ and $\rho_g$
\begin{align}
  \label{Beq:bertin-coeffs}
  \alpha&=\left(c-\frac{v_0^2}{2c}\right)\left(\nu_1\rho_g+\nu_2\right) \qquad \xi=\frac{v_0^3}{2c^2}\nu_1+\tilde\gamma \\
  a_2&=\mu_2\nu_2+(\mu_2\nu_1-\mu_1\nu_2)\rho_g-\mu_1\nu_1\rho_g^2 \nonumber\\
  a_3&=\frac{v_0}{c}\left(2\rho_g\mu_1\nu_1+\mu_1\nu_2-\mu_2\nu_1\right)\nonumber\\
  a_4&=\tilde\zeta-\frac{v_0^2}{c^2}\mu_1\nu_1\nonumber
\end{align}
where $\mu_{1,2}$ are defined by $\mu=\mu_1\rho-\mu_2$ and $\nu_{1,2}$
by $\nu^{-1}=\nu_1\rho+\nu_2$.


\begin{thebibliography}{99}
\bibitem{foot1} In the Vicsek model, the region in which
  solitary bands can be seen vanishes as the system size is taken to
  infinity. At fixed intensive parameters, doubling the system size
  indeed doubles the number of bands and increasing the system size
  hence asymptotically leads, in principle, to a smectic arrangements
  of bands.
\bibitem{foot3} Note that $m$ is actually a momemtum field in this context.
\bibitem{foot2} Again, in the infinite-size limit, the homoclinic
  orbits/solitary wave solutions are only observable in a
  codimension one subset of parameter space.
\end{thebibliography}
\end{document}